\documentclass[runningheads,natbib]{svjour2}
\bibpunct{(}{)}{;}{a}{}{,}

\smartqed

\def\aap{A\&A }
\def\aaps{A\&A Suppl. }
\def\aj{Astron. J. }
\def\apj{ApJ }
\def\apjs{ApJ Suppl.}
\def\apss{Astrophys. Space Sci. }
\def\apjl{ApJ }

\def\mnras{MNRAS }

\usepackage{graphicx}

\journalname{Astronomy and Astrophysics Review}

\begin{document}

\title{Non-thermal emission processes in massive binaries\footnote{The original publication is available at www.springerlink.com (DOI 10.1007/s00159-007-0005-2)}}
\subtitle{}

\titlerunning{Non-thermal emission processes in massive binaries}

\author{Micha\"el De Becker}

\authorrunning{M. De Becker}

\institute{M. De Becker \at
	      Postdoctoral Researcher FNRS, Belgium\\
              Institut d'Astrophysique et G\'eophysique, Universit\'e de Li\`ege, 17 all\'ee du 6 Ao\^ut, Sart-Tilman, 4000, Belgium \\
              Tel.: +32-4-3669717\\
              Fax: +32-4-3669746\\
              \email{Michael.DeBecker@ulg.ac.be}\\}

\date{Received: date}

\maketitle

\begin{abstract}
In this paper, I present a general discussion of several astrophysical processes likely to play a role in the production of non-thermal emission in massive stars, with emphasis on massive binaries. Even though the discussion will start in the radio domain where the non-thermal emission was first detected, the census of physical processes involved in the non-thermal emission from massive stars shows that many spectral domains are concerned, from the radio to the very high energies.

First, the theoretical aspects of the non-thermal emission from early-type stars will be addressed. The main topics that will be discussed are respectively the physics of individual stellar winds and their interaction in binary systems, the acceleration of relativistic electrons, the magnetic field of massive stars, and finally the non-thermal emission processes relevant to the case of massive stars. Second, this general qualitative discussion will be followed by a more quantitative one, devoted to the most probable scenario where non-thermal radio emitters are massive binaries. I will show how several stellar, wind and orbital parameters can be combined in order to make some semi-quantitative predictions on the high-energy counterpart to the non-thermal emission detected in the radio domain.

These theoretical considerations will be followed by a census of results obtained so far, and related to this topic. These results concern the radio, the visible, the X-ray and the $\gamma$-ray domains. Prospects for the very high energy $\gamma$-ray emission from massive stars will also be addressed. Two particularly interesting examples -- one O-type and one Wolf-Rayet binary -- will be considered in details. Finally, strategies for future developments in this field will be discussed.
\keywords{Radiation mechanisms: non-thermal \and Stars: early-type \and Stars: binaries: general \and Radio continuum: stars \and X-rays: stars \and Gamma rays: theory}
\end{abstract}

\section{Introduction \label{intro}}
Stars with masses higher than about 10 solar masses are often referred to as massive -- or early-type -- stars. This category includes O- and early B-type stars, along with Wolf-Rayet (WR) stars which are believed to be the evolved counterparts of O and B stars that have lost a substantial fraction of their mass through their strong stellar winds.

Observational studies of massive stars reveal a wealth of crucial information concerning the physical processes at work in their expanding atmospheres. For instance, early-type stars have been detected in the infrared and radio domains. Independently, \citet{WB} and \citet{PF} developed models to explain the production of the radio and infrared spectra of massive stars: the emission at these wavelengths consists of free-free radiation produced by electrons present in the plasma of their extended envelope. In this emission process, it is thus the stellar wind of the massive star -- and not the star itself -- that is the source of radiation. Moreover, the electrons involved in this process are distributed in energy according to a Maxwell-Boltzmann law: it is therefore a {\it thermal} emission process. The main characteristic of this {\it thermal} radiation is a continuum emission that can be described by a power law of the type:
$$
S_\nu\,\propto\,\nu^\alpha
$$
where $S_\nu$ is the flux density, $\nu$ is the frequency, and $\alpha$ is the spectral index of the free-free emission. For a homogeneous mass loss, at constant or accelerating velocity, $\alpha$ is about equal to 0.6, in agreement with many observations. This thermal emission is intimately related to the mass loss rate of massive stars. A popular method to determine mass loss rates is moreover based on the measurement of the thermal radio flux \citep{WB,PF}.

However, some of the massive stars observed in the radio domain present significant deviations from this well-defined behaviour (see e.g. \citet{WhBe,ABC}, or \citet{rauwHK} for a review). The main characteristics of such a radio emission disagreeing with the classical thermal emission discussed above are the following:
\begin{enumerate}
\item[-] the spectral index ($\alpha$) is not equal to 0.6, but is significantly lower than the thermal spectral index, and might even be negative.
\item[-] the brightness temperature of the radio emission is much higher ($\sim$ 10$^6$ -- 10$^7$\,K) than for thermal emission ($\sim$ 10$^4$\,K).
\item[-] in most cases the radio emission is not steady, and the flux and/or the spectral index can present strong variations.
\end{enumerate}
This {\it non-thermal} radiation was proposed to be synchrotron radiation by \citet{Wh}. Such a radiation mechanism requires two main ingredients. First, the existence of {\it a magnetic field} is needed. The direct detection of the magnetic field of massive stars is not an easy task, mostly because of the difficulty to investigate a Zeeman splitting in line profiles already broadened due to the fast rotation of the star. However, some recent observations led to the direct measurement of the magnetic field strength in the case of a few O- and B-type stars. Second, {\it a population of relativistic electrons} is required. The acceleration of particles in astrophysical sources has mainly been proposed to proceed through the Fermi mechanisn, in the presence of hydrodynamic shocks (see Section\,\ref{accphys}).\\

From this preliminary discussion, it becomes clear that one may discriminate between two main kinds of radio emission, produced by two distinct mechanisms whose physical requirements are very different: one is {\it thermal}, the other one is {\it non-thermal}. In this paper, a census of physical processes likely to lead to the production of non-thermal radiation will be made. The main aspects that I will address are respectively the physics of stellar winds and their interaction in massive binaries, the acceleration of particles, the magnetic field of massive stars, and finally the non-thermal emission processes likely to be at work in the context of early-type stars.

\section{Stellar winds and wind interactions \label{windphys}}

Massive stars are known to produce strong stellar winds. In the case of O-type stars, these stellar winds can reach velocities up to 2000--3000\,km\,s$^{-1}$ with mass loss rates of about 10$^{-6}$--10$^{-5}$\,M$_\odot$\,yr$^{-1}$. The mass loss from massive stars is driven by their strong radiation field, able to transfer momentum to the ions present in their outer envelope. This line driving mechanism was first proposed by \citet{LS} and developed by \citet{CAK} -- the so-called CAK-theory -- before being further investigated and refined for instance by \citet{AL}, \citet{Vink1}, \citet{paul1,paul2} and \citet{kud}. A general discussion of this process can be found in \citet{master}.\\

In the basic CAK-theory, the stellar wind is considered to be stationary, i.e. the radial flow of matter is smooth. However, \citet{LW} pointed out the fact that radiatively driven stellar winds should be unstable, leading to the appearance of perturbations in the mass outflow. These instabilities, generating intrinsic hydrodynamic shocks and clumps in stellar winds,  were investigated in details for instance by \citet{OR}, \citet{OCR} and \citet{Feld}. The presence of the structures resulting from outflow instabilities is believed to produce several observational phenomena:
\begin{enumerate}
\item[-] {\it statistical fluctuations in emission lines in the optical domain}, in the case of O-type (see e.g. \citet{ever1,ever2}) and WR stars (see e.g. \citet{MofWR,BrownWR}). Such fluctuations have generally low amplitudes (of order a few percent at most) and individual `clumps' affect rather narrow velocity intervals.
\item[-] {\it the emission of thermal X-rays}, likely produced by hydrodynamic shocks between plasma shells travelling at different velocities in the wind of isolated stars, and leading to local temperature enhancements of a few 10$^{6}$\,K (see \citet{FeldX}). The thermal nature of the X-rays produced within stellar winds is clearly revealed in their spectra by emission lines associated to highly ionized elements such as O, N, Si, Mg, Ne or Fe. Such emission line spectra have been observed by the {\it XMM-Newton} and {\it Chandra} X-ray observatories \citep{schulz-emline,kahn-emline}.
\end{enumerate}

Beside these intrinsic radiatively driven instabilities, the winds of massive stars can be involved in strong collisions if the star is a member of a massive binary system. In this scenario, the two companions produce their own stellar winds, that interact in a wind-wind collision zone somewhere between the two stars. This interaction zone is confined between two hydrodynamic shocks \citep{SBP,PS}.\\

A consequence of the colliding winds is the production of thermal X-rays, in addition to those produced intrinsically in individual stellar winds. The temperature of the plasma heated by this collision can reach a few 10$^{7}$\,K, while that of individual stellar winds reaches at most a few 10$^{6}$\,K. This additional emission component generally makes binary systems brighter in X-rays than the individual stars constituting them. Empirically, the ratio between the X-ray and bolometric luminosities of a single star\footnote{A canonical relation between X-ray and bolometric luminosities based on a large albeit heterogenous sample of O-stars was proposed by \citet{BSDC}. More recently, \citet{sanalrel} established a similar relation on the basis of a more reduced sample of stars that are members of the open cluster NGC\,6231. The homogeneity of this latter sample led to a dispersion of the canonical relation of only about 40\,\%, whilst that of \citet{BSDC} is about a factor 2.} is about 10$^{-7}$, and any strong excess with respect to this ratio could be considered as an indication of an interaction between stellar winds in a binary system.

\begin{figure}[h]
\centering
\includegraphics[width=7.0cm]{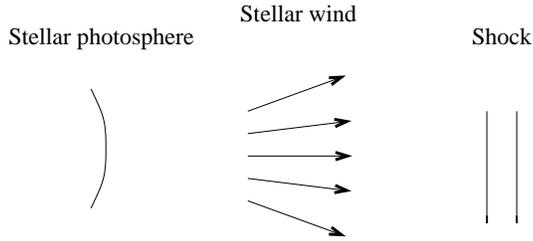}
\caption{Schematic view of the basics of the physics of stellar winds. The wind of a massive star is first produced through radiative driving. The stellar wind gives rise to instabilities that take the form of intrinsic hydrodynamic shocks. In the case of binary systems, the interaction of the stellar winds between the two companions can produce additional shocks in a collision zone.\label{windsum}}
\end{figure}

In summary, as illustrated in Fig.\,\ref{windsum}, hydrodynamic shocks are common features in stellar wind physics, be they due to intrinsic instabilities or to wind-wind collisions in binary systems. This is a crucial point in the context of the non-thermal emission from massive stars, as it appears to be a mandatory ingredient required to accelerate particles.

\section{Particle acceleration mechanism\label{accphys}}

When dealing with the hydrodynamics of stellar winds (individually, or in binary systems), an important concept worth introducing is that of {\it strong shocks}. Strong shocks are particular cases of hydrodynamic shocks where the pressure of the {\it upstream} gas (the gas that has not yet crossed the shock, i.e.\,{\it pre-shock} gas) is negligible in comparison to that of the {\it downstream} gas (the gas that has already travelled through the shock, i.e.\,{\it post-shock} gas). Consequently, if the flow is considered to consist of a perfect gas, we obtain the following relation for the {\it compression ratio ($\chi$)}:
\begin{equation}
\chi\,=\,\frac{\rho_d}{\rho_u}\,=\,\frac{v_u}{v_d}\,=\,\frac{\gamma + 1}{\gamma - 1}
\label{chishock}
\end{equation}
where $\rho_d$ (resp. $v_d$) and $\rho_u$ (resp. $v_u$) are the downstream and upstream gas densities (resp. velocities). For a monoatomic gas, the adiabatic coefficient ($\gamma$) is equal to 5/3, and we derive one of the main properties of strong shocks, i.e. $\chi$\,=\,4.\\

A first approach to explain the acceleration of particles was proposed by \citet{fer}. In his scenario, particles reflected between randomly moving ``magnetic mirrors'' (i.e. irregularities or turbulences in the magnetic field) are accelerated, resulting in a relative increase of their energy ($\Delta$\,E/E) of the order of $(V/c)^2$, where $V$ is the velocity of the shock in the frame of the ambient plasma and $c$ is the speed of light.\\

A variant of this original idea, but in the presence of strong hydrodynamic shocks, was then developed for instance by \citet{BO} and \citet{bella,bellb}. In this context, the crucial point is that, from the particles' point of view, only head-on collisions occur and the accelerated particles do not undergo the decelerating effect of any following collision, contrary to what happens in the original scenario proposed by Fermi. In the literature, this acceleration process is also referred to as the Diffusive Shock Acceleration -- DSA -- mechanism.

\begin{figure}[h]
\centering
\includegraphics[width=9.0cm]{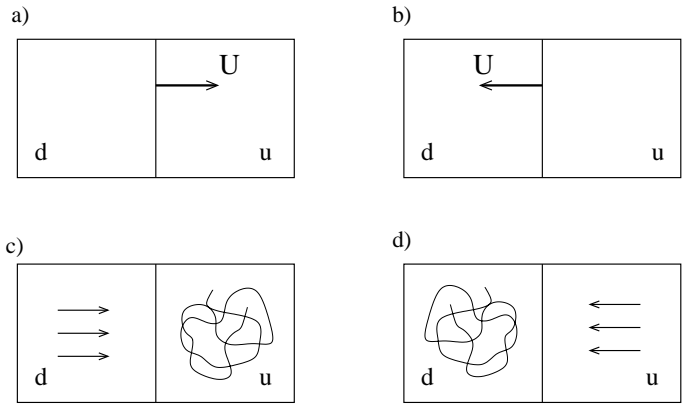}
\caption{Illustration of the dynamics of high-energy particles in the presence of a strong shock where particles are accelerated through the first order Fermi mechanism. The {\it upstream} and {\it downstream} parts of the flow are respectively referred to by the letters $u$ and $d$. The curves in panels (c) and (d) illustrate the scattering due to magneto-hydrodynamic phenomena, and responsible for the isotropic distribution of particles in the frame of reference where the gas is at rest. The four situations are individually described in the text. This figure is inspired from \citet{Long}. \label{fermiframe}}
\end{figure}

This is illustrated in Fig.\,\ref{fermiframe}, where four situations are presented individually. In situation {\bf (a)}, the upstream gas is considered to be stationary, and the shock crosses the medium at a supersonic velocity $U$. If the shock is considered to be stationary ({\bf b}), the upstream gas crosses it with a bulk velocity $v_u\,=\,U$. However, as we are dealing with a strong shock (see Eq.\,\ref{chishock}), the downstream bulk velocity $v_d$ is equal to $v_u/4\,=\,\frac{U}{4}$. In situation {\bf (c)}, i.e. from the point of view of the upstream gas considered to be stationary, the downstream gas moves with a velocity $v_d\prime\,=\,U\,-\,\frac{U}{4}\,=\,\frac{3}{4}\,U$. And in the reverse situation ({\bf d}), i.e. when the downstrem gas is considered to be stationary, $v_u\prime\prime\,=\,\frac{3}{4}\,U$ (see \cite{Long} for details). The main conclusion of this discussion is that the particles crossing the shock encounter gas moving with the same velocity, whatever the direction of the crossing of the shock. Consequently, the particle undergoes exactly the same process of receiving a small amount of energy $\Delta$\,E when crossing the shock from downstream to upstream as it did in travelling from upstream to downstream. There are no crossings in which the particles lose energy : as a result, the process goes linearly with $V/c$, and is therefore much more efficient than the original Fermi mechanism.   

Another crucial aspect of this mechanism is that it is an iterative process. After travelling through the shock front, the high-energy particles can interact with magneto-hydrodynamic waves (Alfv\'en waves) and magnetic turbulence. The consequence is an isotropic scattering of the particles which are allowed to cross the shock once again in the other direction. This results in the multiplication of the crossings, and therefore of the increases in energy, which allows the particles to reach very high energies up to very large Lorentz factors. As a result of this iterative process, the energy spectrum of the relativistic particles is a power law:
\begin{equation}
N(E)\,\propto\,E^{-n}
\label{fermipower}
\end{equation}
where $n$ is the index of the power law. If we are dealing with electrons, this index will be referred to as the {\it electron index} of the relativistic electron population. Following the study of \citet{bella}, the compression ratio of the shock and the index of the relativistic electron population can be related:
\begin{equation}
n\,=\,\frac{\chi + 2}{\chi - 1}
\label{fermiindex}
\end{equation}

\begin{figure}[h]
\centering
\includegraphics[width=6.0cm]{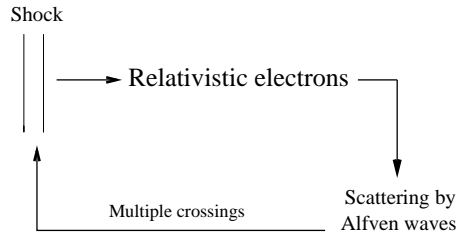}
\caption{Schematic representation of the Fermi acceleration mechanism in the presence of strong shocks. The iterative nature of this process through the interaction of magneto-hydrodynamic phenomena is illustrated. 
\label{fermisum}}
\end{figure}

This general mechanism is believed to be responsible for the production of cosmic rays, probably at least partly accelerated in supernova remnants. In the context of stellar winds of massive stars, the first order Fermi mechanism in the presence of hydrodynamic shocks (intrinsic or in wind-wind interaction zones) is a good candidate to explain the existence of the population of relativistic electrons needed to produce the observed radio synchrotron emission. Figure\,\ref{fermisum} summarizes the iterative process of acceleration of relativistic electrons in the presence of (i) hydrodynamic shocks and (ii) magneto-hydrodynamic phenomena likely to scatter isotropically the high-energy particles. This last point suggests that the role of the magnetic field of massive stars is not negligible in this acceleration process.

Considering the role played by hydrodynamic shocks in the acceleration of particles, two scenarios can thus be envisaged. First, in the {\it single star scenario}, the relativistic electrons are accelerated by intrinsic shocks within stellar winds of isolated stars. Second, in the {\it binary scenario}, colliding winds are required to explain the non-thermal emission detected in the radio domain. To evaluate which scenario is best adapted to explain the non-thermal emission from massive stars, the multiplicity of non-thermal radio emitters must be investigated. At this stage, we know that most of the non-thermal radio emitters -- whatever their spectral type -- are binaries or higher multiplicity systems (see Tables \ref{WRlist} and \ref{Olist}). {\it The binary scenario appears therefore to be the most adapted to explain the non-thermal radio emission from massive stars. In this scenario, the non-thermal emission is produced in the wind-wind interaction zone between the two stars of a binary system, where the relativistic electrons are supposed to be accelerated} \citep{thesis}. Recently, models have been developed applying the DSA mechanism to the particular physical conditions of hydrodynamic shocks of colliding-wind binaries \citep{PD140art,DSACWB}. In the remainder of this paper, I will thus consider that we are dealing with massive binaries.

\section{Magnetic fields of massive stars \label{magnphys}}

The magnetic field of stars is certainly one of the most poorly understood aspects of stellar physics. Observations of the Sun brought a huge amount of relevant information on stellar magnetism, but our general understanding of stellar magnetic fields is still very poor. However, as our knowledge of stellar astrophysics increases, the importance of the role played by magnetic phenomena becomes more and more obvious. In this context, it is worth considering in more detail the generation of magnetic fields within stellar interiors.\\
Section \ref{hc} provides a short discussion of some fundamental hydromagnetic concepts. Sections \ref{fos} and \ref{dt} are devoted to hydrodynamic processes likely to be responsible for the presence of a stellar magnetic field. These discussions are mostly inspired by \citet{Tas} and \citet{Camp}. An alternative approach to produce a magnetic field is briefly described in Sect.\,\ref{alt}. Section\,\ref{magsum} summarizes the various processes possibly at work in the particular case of massive stars. Finally, Sect.\,\ref{meas} gives some words on observational methods used recently to estimate the strength of surface magnetic fields of a few massive stars.

\subsection{Hydromagnetic considerations\label{hc}}

Let us first consider some general equations of electromagnetism. For an electrically conducting fluid such as a plasma, Maxwell's equation relating the curl of the magnetic field ($\textbf{\emph B}$) and the current density ($\textbf{\emph J}$) is the following
\begin{equation}
\nabla\,\times\,\textbf{\emph B}\,=\,\mu_{\circ}\,\textbf{\emph J}
\label{amp}
\end{equation}
where $\mu_{\circ}$ is the permeability of free space. An additional equation useful for our purpose is Faraday's law 
\begin{equation}
\nabla\,\times\,\textbf{\emph E}\,=\,-\,\frac{\partial{\textbf{\emph B}}}{\partial{t}}
\label{far}
\end{equation}
If we include the induction electric field in Ohm's law, we obtain
\begin{equation}
\textbf{\emph J}\,=\,\sigma\,(\textbf{\emph E} + \textbf{\emph v}\,\times\,\textbf{\emph B})
\label{ohm2}
\end{equation}
where $\sigma$, $\textbf{\emph E}$ and $\textbf{\emph v}$ stand respectively for the electric conductivity of the fluid, the intensity of the electric field, and the velocity of the flow.\\
By taking the curl of Eq.\,\ref{ohm2}, we obtain
$$ \nabla\,\times\,\textbf{\emph J}\,=\,\sigma\,\Big[\nabla\,\times\,\textbf{\emph E} + \nabla\,\times\,(\textbf{\emph v}\,\times\,\textbf{\emph B})\Big] $$
Using Eq.\,\ref{amp} in the left hand side and Eq.\,\ref{far} for the first term of the right hand side yields
$$ \nabla\,\times\,\frac{\nabla\,\times\,\textbf{\emph B}}{\mu_{\circ}}\,=\,\sigma\,\Big[-\,\frac{\partial{\textbf{\emph B}}}{\partial{t}} + \nabla\,\times\,(\textbf{\emph v}\,\times\,\textbf{\emph B})\Big] $$
$$ \frac{1}{\mu_{\circ}\,\sigma}\nabla\,\times\,(\nabla\,\times\,\textbf{\emph B})\,=\,-\,\frac{\partial{\textbf{\emph B}}}{\partial{t}} + \nabla\,\times\,(\textbf{\emph v}\,\times\,\textbf{\emph B}) $$
where we considered the permeability of free space as being constant. If we define the magnetic diffusivity as $\lambda\,=\,\frac{1}{\mu_{\circ}\,\sigma}$, and use the vector identity $\nabla\,\times\,(\nabla\,\times\,\textbf{\emph B})\,=\,\nabla\,\cdot\,\nabla\,\cdot\,\textbf{\emph B} - \nabla^2\,\textbf{\emph B}$ where the first term of the second hand side is equal to zero because $\textbf{\emph B}$ is divergenceless, we finally obtain the {\it induction equation}:

\begin{equation}
\frac{\partial{\textbf{\emph B}}}{\partial{t}}\,=\,\nabla\,\times\,(\textbf{\emph v}\,\times\,\textbf{\emph B}) + \lambda\,\nabla^2\,\textbf{\emph B}
\label{ind}
\end{equation}

This equation expresses the temporal variation rate of the magnetic field in a plasma of finite conductivity through two terms: (i) the first one, also called the motion term, corresponding to a gain in intensity, and (ii) the second one standing for the decay of the magnetic field due to the resistivity of the medium \citep{Camp}.

\subsection{Fossil magnetic field\label{fos}}

It has been proposed that the magnetism of some stars could be the relic of a magnetic field that was present in the gas cloud from which the star was formed by gravitational collapse (see e.g. \citet{Tas} for a discussion). When trapped in a star without any mass motion, such a magnetic field would decay due to the finite conductivity of the stellar material. This can be illustrated by the {\it diffusion equation}, which expresses the decay of the magnetic field without any source of magnetic energy.\\

If one neglects induction in Ohm's law, the first term on the right-hand side of Eq.\,\ref{ind} vanishes, and one obtains an estimate of the diffusion time-scale ($\tau_d$) through the following relation
\begin{equation}
\tau_d\,=\,\frac{l^2}{\lambda}
\end{equation}
where $l$ is the typical length scale of the plasma. Consequently, if the plasma is turbulent, $l$ will decrease substantially, leading to a rapid decay of the magnetic field.

\subsubsection{The Sun}

In the case of the Sun, \citet{cow} showed that the time-scale of electromagnetic decay is about 10$^{10}$ years, i.e. the same order of magnitude as its evolution time-scale on the main-sequence. This suggests that any putative primeval magnetic field could possibly be responsible for its current magnetic activity. Consequently, the fossil magnetism has been considered as a viable candidate to explain the magnetism of the Sun, and other stars as well.\\
However, it is not yet clear whether or not a large-scale primeval magnetic field could survive the pre-main-sequence phases. Moreover, in the particular case of the Sun, it is commonly assumed that the magnetic field is continuously generated, and that all traces of primeval magnetism have long since vanished. The reasons are the following: (i) a turbulent convection in the surface layers would cause the decay of the fossil field on a very short time-scale (typically a decade), and (ii) the polar solar magnetic field shows periodic reversals in contradiction with the existence of a stable primeval magnetism\footnote{We note that such a periodic reversal of the stellar magnetic field has also been observed in the case of another star, i.e. the K0 dwarf LQ\,Hydrae \citep{LQ}.}. For these reasons, an alternative theory is needed. This theory, i.e. the {\it hydromagnetic dynamo}, is briefly described in Section\,\ref{dt}.

\subsubsection{Chemically peculiar (CP) stars}

Chemically peculiar (CP) stars are objects displaying atypical abundances, along with abundance inhomogeneities over their surface that may be due to the diffusion of elements under the influence of the stellar magnetic field. The fact that only about 10\,\% of the near-main sequence stars in the same stellar mass range as CP stars display strong magnetic fields (the so-called `10\,\% problem') could possibly be explained by a fossil origin \citep{Moss}. The amount of magnetic flux trapped during star formation is tentatively expected to vary substantially, with only the high flux tail of the distribution resulting in stars with a detected magnetic field on the main sequence. Magnetic fields of a few 10$^{2}$ to several 10$^{4}$\,G have been measured for chemically peculiar Ap and Bp stars \citep{MMC,hub}.

\subsubsection{White dwarfs and neutron stars}

The scenario of the fossil origin of the stellar magnetic field has recently been reconsidered in the context of the study of magnetic fields of compact degenerate stars such as white dwarfs (WD) and neutron stars (NS). More details can be found in \citet{tout} and \citet{FW}.

In this context, a fossil magnetic field is believed to be trapped in stars during their formation process and the magnetic flux should be conserved. For stellar masses lower than about 2\,M$_{\odot}$, the star becomes fully convective during the Hayashi sequence and the fossil magnetic field has no chance to survive. This is e.g. the case of the Sun where the magnetism must be explained in another way (see the dynamo mechanism in Section\,\ref{dt}). For masses larger than about 2\,M$_{\odot}$, the star does not experience a totally convective episode during the Hayashi phase and some fossil magnetic field may survive. The conservation of the magnetic flux as the stellar radius shrinks from that typical for main-sequence OB stars to that of a NS provides an apparently consistent relation between magnetic field strengths during the main-sequence (i.e. a few G to a few kG) and those observed for NSs ($\sim$\,10$^{11}$--10$^{15}$\,G). These numbers seem to lend some support to the conservation of the magnetic flux initially of fossil origin to explain the presence of the magnetic field in massive stars. In this scenario, inhomogeneities in the very weak magnetic field present in extended regions of the Galaxy are likely to produce a large dispersion of stellar magnetic field strengths.

The conservation of the magnetic flux provides a possible interpretation to explain the presence of a magnetic field in WDs as well. In this case, magnetic field strengths of about a few kG \citep{aznar} to 10$^6$--10$^9$\,G \citep{Schm} have been observed. Such a dispersion in the magnetic field strength is also proposed to derive from heterogeneities of the local Galactic magnetic field. 

\subsubsection{Massive non-degenerate stars}
As indicated above, the conservation of the primordial magnetic flux has been proposed to be responsible for the magnetism of massive stars (see e.g. \citet{FW}). However, we can subdivide the inner structure of massive stars into two main zones: (i) the convective core where the fossil magnetic field is unlikely to survive, and (ii) the radiative envelope which results partly from accretion, and which is likely to harbour some residual fossil magnetic field. Moreover, the part of the massive star that shrinks to produce the NS is the core, and not the envelope. So, how can we explain that surface (main-sequence) magnetic fields fit so well core (NS) values after the supernova explosion? Is it fortuitous, or is there any strong physical relation between the convective core and the radiative envelope from the magnetic point of view? At this stage, this question is not answered, and this thus suggests that more than a simple magnetic flux conservation is needed to understand the processes responsible for the observed magnetism of massive stars, be they degenerate or not. Alternative scenarios are discussed in Sections\,\ref{dt} and \ref{alt}.

\subsection{Dynamo theory\label{dt}}

The classical principle of the dynamo action relies essentially on the fact that the motion of a conducting fluid (i.e.\,a plasma) is able to produce electromotive forces capable of maintaining an initial magnetic field. The energy is supplied by the forces driving the fluid flow before being converted into magnetic energy. Consequently, as a first approximation, we will consider that the initial energy supply is huge, and mostly unaffected by this energy conversion process.\\
In order to obtain the dynamo equation, we will use the same formalism as for the induction equation (Eq.\,\ref{ind}). This dynamo equation will then be discussed in the framework of stellar magnetic fields.

\subsubsection{The dynamo equation}

For many astrophysical bodies, $\tau_d$ is shorter than their age, thus they need internal motions of conducting material to explain the observed magnetic fields that would otherwise have undergone ohmic decay down to negligible levels. This issue is investigated in the present section aiming at establishing an equation similar to Eq.\,\ref{ind}, but specific to the case of the dynamo effect\footnote{This discussion is mostly inspired by \citet{Tas} and \citet{Camp}.}.\\

Let's consider an axisymmetric magnetic field that can be split into poloidal and toroidal components. According to `Cowling's theorem', {\it a steady axisymmetric magnetic field cannot be maintained by dynamo action} \citep{cow}. The problem is that in astrophysical bodies, the symmetry is dominantly axial. A solution to solve this problem was proposed by \citet{Par}, who noted that convection can cause rising cells of fluid to rotate and carry lines of force of the toroidal field into loops. Consequently, a large number of such loops could regenerate the poloidal component of the magnetic field. Such a turbulence is expected to arise from the Coriolis force, and is most commonly called the {\it $\alpha$-effect}.\\

The velocity and magnetic fields can be expressed as a sum of large-scale (mean field) and small-scale (fluctuating field) contributions. The $\alpha$-effect is a small-scale turbulence acting on the mean (large-scale) magnetic field through a mean electromotive force ($\epsilon$) given by the following relation
\begin{equation}
{\bf \epsilon}\,=\,\alpha\,\textbf{\emph B} - \beta\,\nabla\,\times\,\textbf{\emph B}
\end{equation}
where $\alpha$ is a coefficient that vanishes for homogeneous isotropic turbulence, and $\beta$ can be regarded as an additional contribution to the fluid's magnetic diffusivity due to the turbulence (also called eddy diffusivity).\\

Taking into account this additional electric field in the induction equation, we obtain the following equation for the mean field, i.e. the {\it dynamo equation}
$$
\frac{\partial{\textbf{\emph B}}}{\partial{t}}\,=\,\nabla\,\times\,(\textbf{\emph v}\,\times\,\textbf{\emph B}) + \nabla\,\times\,(\alpha\,\textbf{\emph B}) + (\lambda + \beta)\,\nabla^2\,\textbf{\emph B}
$$
This equation allows axisymmetric solutions to exist for the mean magnetic field. As the magnetic diffusivity is essentially dominated by the eddy diffusivity, it is often neglected in the formulation of the dynamo effect.\\
In this last relation, the magnetic and eddy diffusivities have been considered to be constant throughout the plasma. However, a more general formulation of the {\it dynamo equation} would be

\begin{equation}
\frac{\partial{\textbf{\emph B}}}{\partial{t}}\,=\,\nabla\,\times\,(\textbf{\emph v}\,\times\,\textbf{\emph B}) + \nabla\,\times\,(\alpha\,\textbf{\emph B}) - \nabla\,\times\,\Big[(\lambda + \beta)\,\nabla\,\times\,\textbf{\emph B}\Big]
\label{dyn}
\end{equation}

With appropriate boundary conditions, this dynamo problem turns out to be an eigenvalue problem, aiming at finding eigenvalues ($\xi$) corresponding to non-decaying solutions of the type $\textbf{\emph B(x,t)}\,=\,\textbf{\emph B(x)}\,\exp(\xi\,t)$, where $x$ stands for the spatial position vector. The study of \citet{CM} shows that the solutions for massive main-sequence stars (convective core + radiative envelope) point to magnetic fields that are concentrated in the immediate vicinity of the core-envelope interface. This is interpreted as a consequence of the fact that the magnetic diffusivity is expected to decrease rapidly as one moves from the turbulent core to the stably stratified radiative envelope.\\

The expression of the velocity can contain a term accounting for the differential rotation. In that case, the expressions for the poloidal and toroidal components of the magnetic field show that a magnetic field can be produced through two main mechanisms, respectively called the $\alpha$- and $\Omega$-effect:
\begin{enumerate}
\item[] {\bf $\alpha$-effect.} As stated above, it is the production of a magnetic field from turbulence arising from the Coriolis force. This process is able to produce both toroidal and poloidal fields. 
\item[] {\bf $\Omega$-effect.} It consists in the production of a toriodal field through the shearing of the poloidal component of the field due to the non-uniform rotation of the star.
\end{enumerate}
The general case where both effects act simultaneously is often called the $\alpha^2\Omega$-dynamo. Particular dynamo cases can also be considered, such as the $\alpha^2$-dynamo where the $\Omega$-effect can be neglected, or the $\alpha\Omega$-dynamo where the impact of the $\alpha$-effect can be neglected with respect to that of the $\Omega$-effect in the production of a toroidal field from the poloidal field.\\

Stars with very low mass (M\,$\leq$\,0.4\,M$_\odot$) are essentially fully convective. The dynamo process can occur within the whole stellar interior. 
In the case of stars of solar and later types (M\,$\leq$\,1\,M$_\odot$), the hydromagnetic dynamo is believed to occur somewhere within the outer convective zone. For instance, a magnetic field of a few Gauss is observed over most of the surface of the Sun, with strong localized fields of up to 4000 G in sunspots (see e.g. \citet{Tas}). In this case, cyclic variations of the chromospheric activity suggest that the Sun-like dynamo produces a time-dependent field responsible for a cyclic magnetic activity.

For massive stars (M\,$\geq$\,8-10\,M$_\odot$), the study of \citet{CM} shows that such a dynamo action is possible in the core. However, the question is how a magnetic field generated by a core dynamo could travel up to the surface of the star where it can be detected directly (e.g. using spectropolarimetric methods, see Sect.\,\ref{meas}) or indirectly (e.g. synchrotron radio emission, putative magnetic stellar wind confinement). Two scenarios have currently been proposed to address this issue, and they are briefly discussed below.

\subsubsection{How to bring the magnetic field to the surface?\label{transp}}

The transport of the magnetic field of massive stars from the convective core to the surface constitutes a challenge for the stellar magnetism theory. Two scenarios have been proposed: (i) the meridional circulation through Eddington-Sweet cells \citep{CM}, and (ii) the buoyant rise of magnetic tubes along the rotation axis \citep{MC}.\\

\paragraph{Meridional circulation -- Eddington-Sweet cells.}
The rotation of early-type stars can lead to departures from sphericity, responsible for pole-equator temperature differences in the radiative interior (see e.g. \citet{CM}). These temperature differences cannot be balanced under the assumption of simultaneous radiative and hydrostatic equilibria. The reaction of the star is to generate a large-scale circulation in the meridional plane, known as the Eddington-Sweet circulation. As some of these flow streamlines come close to the surface of the star, and also close to the boundary between the convective core and the radiative envelope, Eddington-Sweet cells may be able to bring the magnetic field from the core up to the surface.\\
According to the study of \citet{CM}, in the case of models with a strong magnetic diffusivity contrast between the core and the envelope (that are probably the most realistic ones), the dynamo effect is inhibited by the meridional circulation if it reaches a level suitable for a significant magnetic flux transport. Moreover, such meridional circulation regimes are not likely to be attained in the interior of early-type main-sequence stars. Consequently, an alternative scenario is needed to carry magnetic fields from the stellar core to the surface.\\

\paragraph{Buoyant rise of magnetic tubes.}
In this model of advection of magnetic flux, \citet{MC} consider the dynamics of a circular ring, symmetric about the stellar rotation axis. This ring is an isolated concentration of magnetic field, which is allowed to begin an outward motion from the convective core thanks to the buoyancy. In this model, the ring, or magnetic tube, is in mechanical equilibrium with the surrounding medium. We note however that the radiation pressure is neglected in such a simplified approach, therefore restricting the discussion to stars with masses up to about 10 M$_\odot$. The rotation is considered to be homogeneous, and no meridional circulation is taken into account.\\
The time-scale of the rise of the magnetic tubes is shorter than the typical evolution time-scale of massive stars on the main-sequence. However, the density deficit of such a ring is expected to decrease as it travels through the radiative envelope, causing the upward buoyant acceleration experienced by the ring to decrease as well. The ring indeed spends considerably more time (about a factor 100) travelling through a thin shell between 0.90 and 0.96 R$_\star$, than in ascending from the core to a radius of 0.90 R$_\star$. The inclusion of meridional circulation is expected to significantly decrease the time needed to transport the magnetic flux through the layers just below the surface. The joint action of the buoyancy transport and of the meridional circulation could therefore possibly lead to short rise times, able to transport a significant magnetic flux to the surface during the very first stages of the main-sequence life of massive stars.

\subsection{An alternative dynamo scenario\label{alt}}
A completely different possibility worth considering is the production of a magnetic field through a dynamo mechanism just below the surface, without the need of bringing it from the core. This approach is described for instance by \citet{spr2} or by \citet{MacM}, showing that a dynamo can be driven by the differential rotation in a stable stratified zone.\\

According to the study of \citet{spr2}, an instability of an initial toroidal field (even small) creates a vertical field component. Such an instability, called Tayler instability \citep{spr1}, consists of non-axisymmetric horizontal displacements of fluid elements. This vertical component of the field is distorted by differential stellar rotation which converts it into additional toroidal field lines. These horizontal field lines become progressively closer and denser, therefore producing a much stronger horizontal field.\\

This mechanism is usually refered to as the Tayler-Spruit dynamo \citep{spr2,MM1,MM2}. In the context of massive stars, this scenario could be envisaged within the radiative envelope, therefore producing a magnetic field close to the surface.

\subsection{Summary\label{magsum}}
Considering the scenarios discussed in this section, we see that the origin of the magnetic field in massive stars is not yet well established. Still, we can make a census of the physical processes likely to play a role in this context.

A fossil magnetic field is unlikely to survive in the turbulent regime of the core of the massive star, but a classical dynamo action might produce a magnetic field in that part of the star. Such a classical dynamo cannot be at work in the radiative envelope. However, an alternative dynamo driven by the differential rotation of the star may produce a magnetic field close to the surface. Moreover, the possibility to bring to the surface some magnetic field produced in the core thanks to the combined action of meridional circulation and buoyancy cannot be completely rejected. We also note that a contribution of a fossil magnetic field might be present in the radiative envelope, as it results partly from the accretion of interstellar material that has not experienced a fully turbulent episode.

\subsection{Direct determination of magnetic field strengths of massive stars\label{meas}}

In the presence of a magnetic field, the degeneracy of atomic sublevels is lifted, splitting a given spectral line into several components (Zeeman effect). This effect has also the particularity of polarizing the light emitted through radiative transitions between sublevels affected by this splitting. The Zeeman effect is mostly efficient in circular polarization. Among the Stokes parameters characterizing the state of polarization of quasi-monochromatic light, this circular component corresponds to the Stokes V parameter and responds to the line-of-sight component of the magnetic field (longitudinal Zeeman effect). It indicates the intensity difference between right- and left-handed circular polarizations. The other Stokes parameters respectively measure the unpolarized (I) and linearly polarized (Q and U) light.\\
In the case of massive stars, the strong rotational broadening due to the large projected rotational velocities (V$_{rot}\,\sin\,i\,\approx$\,100 - 450 km\,s$^{-1}$) does not allow to detect directly the line splitting (or broadening) due to the Zeeman effect. Nevertheless, the Zeeman polarization can be observed through spectropolarimetric methods. Recently, on the basis of this principle, the intensity of the magnetic fields of only a few massive stars have been estimated. Three examples are respectively $\beta$\,Cep (B1IV), $\theta^1$\,Ori\,C (O4-6V), and $\zeta$\,Cas (B2IV). With this method, for a simple dipolar magnetic field model, B$_p$ and $\theta$ for $\beta$\,Cep are respectively of 360 $\pm$ 60 G and 85$^\circ$ $\pm$ 10$^\circ$ \citep{Don2}. The same parameters for $\theta^1$\,Ori\,C are of 1100 $\pm$ 100 G and 42$^\circ$ $\pm$ 6$^\circ$ \citep{Don3}, and 335$_{-65}^{+120}$\,G and 18$^\circ$ $\pm$ 4$^\circ$ for $\zeta$\,Cas \citep{Nei}. In the case of these stars, more sophisticated models including higher order field components (quadrupolar,...) donot improve significantly the fit. Finally, we note that as this technique relies on a particular Zeeman component (the longitudinal one), it is sensitive to the phase of the rotational cycle of the star. Consequently, observing exposure times must be significantly shorter than the rotational period. For this reason, and also because of the low amplitude of the Stokes signature, this method is only applicable to bright stars\footnote{The method was therefore not applicable to the cases of the stars discussed in Sect.\,\ref{obsres}, though a direct determination of their surface magnetic field strength would be crucial.}. In the future, observations with spectropolarimeters suchs as ESPaDOns \citep{espadons} mounted on the Canada-France-Hawaii Telescope (CFHT) are expected to improve significantly our knowledge of the strength of the magnetic field of massive stars.

\subsection{The magnetic field in the wind interaction zone of massive binaries}

If we consider that there is no magnetic field production in the collision zone of massive binaries, the local magnetic field -- in the interaction zone -- should come from the stars themselves. We need therefore to estimate the dependence of the magnetic field strength as a function of the distance to the star where it is produced. Some equations are given in Section\,\ref{magnfield} to deal with this dependence. In addition, we should consider that two stars may contribute to the local magnetic field and that the magnetic field may undergo some intensification in the post-shock plasma. This latter point has been addressed by \citet{bell2004}, who considered the amplification of the background magnetic field by the accelerated particles. Only a very simplified approach to estimate the strength of the magnetic fied in the colliding wind region is followed in Section\,\ref{magnfield}.

\section{Non-thermal emission processes\label{radphys}}

A crucial point in the context of this paper is the distinction between {\it thermal} and {\it non-thermal} emission processes. In {\it thermal} processes, the energy distribution of the electrons involved in the photon production mechanism (e.g. thermal free-free emission, radiative recombination and collisional excitation followed by radiative decay) can be described by a Maxwell-Boltzmann law. In the case of {\it non-thermal} processes, particles (electrons, protons...) are relativistic and are distributed according to a power law (see Eq.\,\ref{fermipower}). 

Among the non-thermal emission processes, we can distinguish {\it particle-field} interactions (particles interacting with radiation or a magnetic field) and {\it particle-matter} interactions. The most common processes are briefly described below.

\subsection{Particle-field interactions}

\subsubsection{Synchrotron radiation\label{synchradproc}}
A relativistic electron travelling in a magnetic field will be subjected to a Lorentz force orthogonal to the magnetic field direction. Consequently, the electron will adopt a helical path, and will emit photons ({\it synchrotron radiation}).

The characteristic energy of the photons is given by
\begin{equation}
E_c\,=\,\frac{3\,h}{4\,\pi}\,\frac{e\,B\,\sin\,\theta}{m_e\,c}\,\Big[\frac{E}{m_e\,c^2}\Big]^2
\label{ensynch}
\end{equation}
where $h$ is Planck's constant, $e$ the electron charge, $B$ the magnetic field strength, $\theta$ the pitch angle, $m_e$ the mass of the electron, $c$ the speed of light, and $E$ the energy of the electron \citep{CR}. This relation shows clearly that very energetic particles and strong magnetic fields are needed to produce synchrotron radiation in the very high energy domain. However, with rather modest magnetic fields and ``moderate'' relativistic electrons, synchrotron radiation can easily be produced in the radio domain. In the context of massive stars, the non-thermal radio emission discussed in the introduction of this work is most probably produced through this mechanism \citep{Wh}.

It should be noted that the so-called Razin-Tsytovich effect (or Razin effect, see e.g. \citet{sventhesis}) is likely to occur, leading to an apparent suppression of the synchrotron radiation. This effect is based on a particular property of the synchrotron emission: it is characterized by a significant beaming, i.e.\,the radiation is emitted in preferential directions. This property of the synchrotron emission is important considering the fact that, in a plasma such as that constituting stellar winds, the beaming effect can be substantially reduced, leading therefore to a decrease of the synchrotron emitting power in a given direction. This suppression is a characteristic effect of plasma on radiation by relativistic particles. It can be shown that the Razin-Tsytovich effect is a consequence of the suppression of radiation with a frequency below the plasma frequency, after transformation from the rest frame of the relativistic particle to the rest frame of the observer. A detailed discussion on this transformation can be shown in \citet{Melrose-RT}. This effect may be responsible for significant underestimates of the radio synchrotron power of astronomical sources. 

\subsubsection{Inverse Compton scattering}
Particles (electrons or protons) can also interact with photons. In the context of this study, only electrons will be considered. If the typical energy of the relativistic electrons is high enough, the latter can transfer some of their energy to soft photons. This process is known as {\it inverse Compton (IC) scattering}. In this way, moderately relativistic electrons are able to ``promote'' ultra-violet photons to the high-energy domain (X-rays and soft $\gamma$-rays). The relation between the average energy of the emerging photons ($E_{IC}$) and that of the scattered soft photons ($E_{soft}$) is
\begin{equation}
E_{IC}\,=\,\frac{4}{3}\,E_{soft}\,\gamma^2
\label{convic}
\end{equation}
where $\gamma$ is the Lorentz factor characterizing the relativistic electrons. This latter relation is only valid in the Thomson limit. In this regime, the square root of the product of the energies of the interacting electron and photon is significantly lower than the rest energy of the electron \citep{CR}.\\

Most frequently, inverse Compton scattering is considered in the Thomson regime, and in that case the cross-section of the process is the Thomson cross-section ($\sigma_T$, see Sect.\,\ref{heco}). However, when we are dealing with very high-energy particles, a quantum relativistic cross-section has to be used and the process is considered in the Klein-Nishina regime \citep{Long}. As shown by \citet{DSACWB}, the Klein-Nishina cross-section can be significantly lower than the Thomson cross-section in the very high-energy domain. The main effect is a strong decrease in the efficiency of IC scattering to produce very high-energy photons.

IC scattering is of particular interest in the framework of massive stars because of the strong supply of UV and visible photons provided by the photosphere. This process is indeed believed to play a significant role in the cooling of relativistic electrons in stellar wind environments.

\subsection{Particle-matter interactions}

\subsubsection{Relativistic Bremsstrahlung}
When a relativistic electron is accelerated in the electric field of a nucleus or other charged particles, radiation is produced: {\it relativistic Bremsstrahlung}. In general, only the emission from the electron has to be considered. It comes from the fact that the intensity of Bremsstrahlung by a nucleus of mass M is $\sim$\, (M/m$_{e}$)$^{-2}$ times the effect generated by the electron.

We note that such a mechanism requires rather high plasma densities to become significant. For this reason, relativistic Bremsstrahlung is not expected to be responsible for non-thermal emission in the case of O-type stars. This scenario was nevertheless considered by \citet{Pol} in the case of the denser winds of Wolf-Rayet stars. 

\subsubsection{$\pi^0$-decays from proton-proton emission}

Even though we are mostly dealing with electrons in the context of this paper, we should not neglect the fact that protons are very abundant in astrophysical environments such as stellar winds. Relativistic protons can interact with other `ordinary' protons to produce neutral pions ($\pi^0$), that decay into gamma-rays with a proper lifetime of only 9\,$\times$\,10$^{-17}$\,s. For a more detailed discussion of this process I refer to \citet{CR}.\\

It has been proposed that the $\pi^0$-decay mechanism could be at work in young open clusters \citep{Manch}. In this context, it is suggested that the combined winds of massive stars may interact with the gas in the cluster and produce a system of bow shocks. These shocks could then be responsible for the acceleration of protons up to relativistic velocities. Consequently, the relativistic protons may interact to produce a diffuse gamma-ray emission in open clusters and OB associations. In addition, it is important to note that the $\pi^0$-decay mechanism may certainly be responsible for the very high-energy emission discussed later in this review.

\section{The non-thermal high-energy counterpart of a massive binary\label{hent}}

\begin{figure}[h]
\centering
\includegraphics[width=11.5cm]{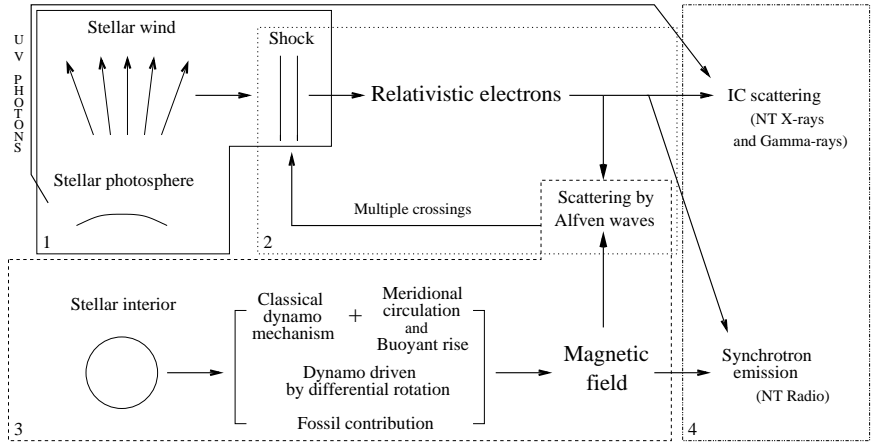}
\caption{Overview of the physical processes likely to contribute to the production of non-thermal emission from massive stars. The main aspects of the problem are subdivided in boxes directly related to the processes discussed with more details in this paper: (1) stellar wind physics (Sect.\,\ref{windphys}); (2) particle acceleration mechanism (Sect.\,\ref{accphys}); magnetic processes (Sect.\,\ref{magnphys}); non-thermal emission processes (Sect.\,\ref{radphys}).\label{overview}}
\end{figure}

I have discussed several aspects of the physics of massive stars, or different topics of astrophysics, likely to play a role in the production of non-thermal emission of massive stars. Figure\,\ref{overview} presents an overview of these physical processes. This schematic view is mostly inspired from that given by \citet{HK}. In this figure, the four boxes individually represent the topics independently discussed in the four previous sections of this chapter in their order of appearance. The intersections between boxes illustrate the noticeable interconnections between these different fields of astrophysics. 

As shown in Sect.\,\ref{radphys}, non-thermal emission processes can also be at work in the high-energy domain. In order to investigate the issue of the high-energy counterpart to the non-thermal emission of massive stars detected in the radio domain, it is helpful to adopt a somewhat more quantitative approach of the physical processes believed to be responsible for the non-thermal emission of radiation in the particular case of binary systems. We note that the description below should be considered as a simplified approach, used only to give an idea of the impact of several physical parameters on some quantities such as the IC luminosity. A more complete and quantitative approach would require sophisticated models including more physics than used below.\\

The radio synchrotron luminosity and the inverse Compton luminosity can be related by a unique equation, provided both emission mechanisms arise from the same population of relativistic electrons \citep{WC}. Such a relation derives from the fact that, for both processes, the total energy lost by a high-energy electron can be expressed by similar relations depending on the energy density respectively of the magnetic field ($U_B$) and of the radiation field ($U_{ph}$). This leads to the following equation for the ratio between the radio synchrotron luminosity ($L_{synch}$) and the high-energy inverse Compton luminosity ($L_{IC}$):
\begin{equation}
\label{lratio}
\frac{L_{synch}}{L_{IC}}\,=\,\frac{U_B}{U_{ph}}
\end{equation} 
We can use this luminosity ratio to obtain an expression of the IC luminosity as a function of the synchrotron radio luminosity, taking into account the fact that both stars of the binary system contribute to the photon energy density:
\begin{equation}
\label{lic}
L_{IC}\,=\,\frac{2}{c}\,\frac{L_{synch}}{B_{col}^2}\,\Big(\frac{L_{bol,1}}{r_1^2} + \frac{L_{bol,2}}{r_2^2}\Big)
\end{equation}
where $L_{bol,1}$ (respectively $L_{bol,2}$) and $r_1$ (resp. $r_2$) stand for the bolometric luminosity and for the distance from the center of the star to the collision zone for the primary (resp. the secondary). $B_{col}$ is the strength of the magnetic field in the acceleration region of relativistic particles (which is also the emitting region of the synchrotron and IC photons, i.e. the collision zone).

This approach is very similar to that of \citet{EU} and \citet{BR}. However, these latter studies took into account only one term for the photon density. This approximation was valid because the authors had to deal with asymmetric binary systems where the winds collide close to the secondary. Consequently, the primary photon density could be neglected compared to that of the secondary. Eq.\,\ref{lic} is therefore more general than that used by \citet{EU} and \citet{BR}, and it should preferentially be used each time one has to deal with binary systems characterized by a momentum ratio ($\eta$) that is not significantly smaller than 1 (see Eq.\,\ref{eta}). It should be mentioned that the simplified approach described below does not take into account the variation of the $U_{B}/U_{ph}$ ratio accross the wind collision region, even though we know that the non-thermal emission region can be extended (see Sect.\,\ref{radiodomain}). A more consistent approach would require a detailed knowledge of the spatial distribution of relativistic particles -- along with that of the magnetic field -- in the wind collision region. In the remaining of this section, I will discuss in detail three points that need to be addressed properly if one wants to apply Eq.\,\ref{lic}: the geometry of the system, the intensity of the local magnetic field, and the synchrotron radio luminosity.

\subsection{Geometry of the system\label{geom}}

\subsubsection{Separation between the stars. \label{sep}}
In an eccentric binary system, the relative distance between the two stars in units of the semi-major axis of the orbit, is given by
\begin{equation}
\label{relsep}
r\,=\,\frac{1 - e^2}{1 + e\,\cos{v}}
\end{equation}
where $v$ is the true anomaly, i.e. the angle between the vector radius and the direction to periastron, and $e$ is the eccentricity of the orbit.
In order to obtain the relative separation as a function of the orbital phase ($\phi$), a relation between the true anomaly and the orbital phase is needed. As a first step, let us consider the expression of the true anomaly ($v$) as a function of the eccentric anomaly ($E$).

\begin{equation}
\label{eccanom}
v\,=\,2\,\arctan\,\Big[\Big(\frac{1 + e}{1 - e}\Big)^\frac{1}{2}\,\tan\,\frac{E}{2}\Big]
\end{equation}

The eccentric anomaly is related to the mean anomaly ($M\,=\,2\,\pi\,\phi$ when $\phi\,=\,0$ at periastron) through Kepler's equation:

\begin{equation}
\label{meananom}
E - e\,\sin\,E\,=\,M
\end{equation}

The numerical resolution of Eq.\,\ref{meananom} allows to obtain the value of $E$ for a given orbital phase. Then the successive application of Eqs.\,\ref{eccanom} and \ref{relsep} provides the relative separation between the two components of the binary system.

The projected semi-major axes ($a_1\,\sin\,i$ and $a_2\,\sin\,i$) for the two components can be obtained from the spectroscopic orbital solution of the binary system. Provided an estimate of the inclination angle $i$ is available, the multiplication of the relative separation ($r$) by the sum $a_1 + a_2$ yields the absolute separation ($D$) between the two stars.

\subsubsection{Wind momentum equilibrium. \label{wmr}}
The collision between the stellar winds of the two components of a binary system occurs at a position which is set by the relative strengths of the winds. If the winds have similar strengths, the collision zone is a plane perpendicular to the line of centres, at equal distance from the two stars. On the contrary, if the winds are not equal, the collision zone is a curved layer of shocked gas partially folded around the star producing the weaker wind.

If we assume that the spherical winds of the two components flow nearly radially out to the shocks, we can derive the distances from the stars where these winds meet. At the position of the collision zone, the wind momenta are balanced, yielding 
\begin{equation}
\label{mom}
r_1\,=\,\frac{1}{1 + \eta^\frac{1}{2}}\,D\,,\,r_2\,=\,\frac{\eta^\frac{1}{2}}{1 + \eta^\frac{1}{2}}\,D\
\end{equation}  
where $r_1$, $r_2$, $D$ and $\eta$ are respectively the distances to the collision zone from the primary and the secondary, the separation between the stars (see above), and a dimensionless parameter expressed as
\begin{equation}
\label{eta}
\eta\,=\,\frac{\dot M_2\,V_{\infty,2}}{\dot M_1\,V_{\infty,1}}
\end{equation}
where $\dot M$ and $V_\infty$ stand for the mass loss rate and the terminal velocity respectively. The parameter $\eta$ thus expresses the ratio of the momentum flux of the two winds \citep{SBP}.

\subsection{Local magnetic field.\label{magnfield}}
The surface magnetic field strength ($B_s$) is probably one of the most critical and poorly known physical quantities for massive stars. Moreover, provided that $B_s$ is determined, the behaviour of the magnetic field as a function of the distance to the star in the ionized material of the stellar wind still needs to be established.

Some relations providing the dependence of the strength and geometry of the magnetic field as a function of the radial distance ($r$) to the star in the outflowing gas are given for instance by \citet{UM}. Three regimes can be envisaged: (1) {\it dipolar}, (2) {\it radial}, and (3) {\it toroidal}. The equations for these three regimes, along with their conditions of application, are given below:
\begin{equation}
\label{magn}
B\,=\,B_s\,\times\,\left\{
\begin{array}{lll}
\Big(\frac{R}{r}\Big)^3	& \,\rm{for}\, R\,\leq\,r\,<\,r_A & (\mathrm{dipolar})\\
 & & \\
\frac{R^3}{r_A\,r^2}	& \,\rm{for}\, r_A\,<\,r\,<\,R\,\frac{V_\infty}{V_{rot}} & (\mathrm{radial})\\
 & & \\
\frac{V_\infty}{V_{rot}}\,\frac{R^2}{r_A\,r}	& \,\rm{fo}r\, R\,\frac{V_\infty}{V_{rot}}\,<\,r & (\mathrm{toroidal})
\end{array}\right.
\end{equation}

In these relations, $V_{rot}$ and $V_\infty$ are respectively the stellar rotational velocity and the wind terminal velocity. $R$ is the stellar radius, and $r_A$ is the Alfv\'en radius defined as the radius where the kinetic and magnetic energy densities are balanced. The Alfv\'en radius can be expressed as follows:

\begin{equation}
\label{alf}
r_A\,\sim\,R\,\times\,\left\{\begin{array}{ll}
		1 + \xi & for\,\xi\,\ll\,1,\\
		\xi^{1/4} & for\,\xi\,\gg\,1
	    \end{array}\right.
\end{equation}
with the parameter $\xi$ expressed by
\begin{equation}
\label{xi}
\xi\,=\,\frac{B_s^2\,R^2}{2\,\dot M\,V_\infty}
\end{equation}
From these relations, we can see that the typical Alfv\'en radius for early-type stars is in the range from $\sim\,R$ to $\sim\,2-3\,R$ \citep{UM}. The best approach consists in using the well-established or expected wind and stellar parameters to estimate which regime is suitable, before applying the corresponding relation to estimate the local magnetic field strength at a given distance $r$ from the center of the star.

When the strength of the magnetic field at the distance of the shock zone is obtained from Eq.\,\ref{magn}, one still has to wonder whether it is affected by the hydrodynamic conditions close to the shock. In the case of a strong shock, the compression ratio ($\chi$) is expected to be equal to 4 (see Sect.\,\ref{windphys}). The magnetic field strength dependence on the density could be of the order of $\rho^{2/3}$ \citep{Tas}, and can be used to determine some kind of concentration factor $\epsilon$. Consequently, the magnetic field strength could increase by up to a factor of 2.5 in the post-shock region of a strong shock. Moreover, the magnetic fields from both stars are expected to contribute to the local magnetic field strength at the collision zone. Let us therefore consider that the mean local magnetic field can be obtained using the following relation:
\begin{equation}
\label{blocal}
B_{col}\,=\,\epsilon\,\frac{B_1 + B_2}{2}
\end{equation}
where $B_1$ and B$_2$ are the local magnetic fields from the primary and the secondary respectively, expressed in Gauss. This approach is thus significantly different from that of \citet{EU} and \citet{BR} where the local magnetic field was considered to come from the secondary star only, as those authors dealt with cases where the collision zone was very close to the secondary. I emphasize on the fact that the way we deal here with the magnetic field is only an assumption. A detailed investigation of the magneto-hydrodynamic conditions at the collision zone is far beyond the scope of the present discussion.

\subsection{Radio luminosity. \label{ralum}}

In the literature, radio emission levels are mostly provided as flux densities expressed in mJy at a given frequency ($S_\nu$). This flux density can be converted into more conventional units (cgs) following the procedure described below.

We first need to establish a relation yielding the flux in a given frequency interval. This can be done through an integration of the flux density between two frequencies $\nu_1$ and $\nu_2$:
\begin{equation}
\label{intradio}
S\,=\,\int_{\nu_1}^{\nu_2}\,S_\nu\,d\nu
\end{equation}
The lower boundary $\nu_1$ is defined by the rather steep increase of the radio flux as a function of frequency likely to occur at frequencies of the order of a fraction of GHz (i.e. at a wavelength of a few tens of cm), before reaching a maximum at a frequency of the order of the GHz. This turnover may be due to the combined action of the opacity of the plasma that decreases with increasing frequency, and of the Razin-Tsytovich effect. Recent modelling and observations seem to favor free-free opacity as the dominant factor at least in the case of WR\,140 \citep{cygint}. For the upper limit on the frequency ($\nu_2$), a value of the order of several tens of GHz (i.e. a fraction of cm) should be considered (see e.g. Sect.\,2.3.1 in \citet{Pit2}). This upper boundary is determined by the maximum energy reached by relativistic electrons that is severely affected by IC cooling (maximum Lorentz factors for electrons of the order of 10$^4$-10$^5$ are expected depending of the colliding-wind binary system, see Sect.,\ref{heco}). It should be noted that these characteristic frequencies depend strongly on the target that is considered. A few examples of synchrotron radio spectra of long period massive binaries can be found for instance in \citet{PD140art} and \citet{Doug}. 
As we know that the radio spectrum is a power law with a spectral index $\alpha$, we can use the following relation:
$$
S_\nu\,=\,S_{\nu'}\,\Big(\frac{\nu}{\nu'}\Big)^\alpha
$$
where $\nu'$ is a frequency for which we have a measure of the flux density. Inserting this relation in Eq.\,\ref{intradio}, we obtain
$$
S\,=\,\frac{S_{\nu'}}{\nu'^\alpha}\,\int_{\nu_1}^{\nu_2}\,\nu^\alpha\,d\nu
$$
and we derive the integrated flux:
\begin{equation}
\label{intradio2}
S\,=\,\frac{S_{\nu'}}{(\alpha + 1)\,\nu'^\alpha}\,\big(\nu_2^{\alpha + 1} - \nu_1^{\alpha + 1}\big)
\end{equation}
As a final step, we consider that
$$ 1\,\mathrm{mJy}=10^{-29}\,\mathrm{W}\,\mathrm{m}^{-2}\mathrm{Hz}^{-1}=10^{-26}\,\mathrm{erg}\,\mathrm{s}^{-1}\mathrm{cm}^{-2}\mathrm{Hz}^{-1}
$$
and we multiply both sides of Eq.\,\ref{intradio2} by $4\,\pi\,d^2$ in order to obtain a relation giving the radio luminosity in cgs units integrated between two frequencies $\nu_1$ and $\nu_2$:
\begin{equation}
\label{lradio}
L_{radio}\,=\,4\,\pi\,d^2\,10^{-26}\,\frac{S_{\nu'}}{(\alpha + 1)\,\nu'^\alpha}\,\big(\nu_2^{\alpha + 1} - \nu_1^{\alpha + 1}\big)
\end{equation}
where $S_{\nu'}$, $\nu'$ and $d$ are respectively expressed in mJy, in Hz and in cm. To calculate this quantity, the flux density at a given frequency ($\nu'$), the distance to the star, and the spectral index are thus needed.\\

In a binary system producing non-thermal radio emission, the total radio spectrum has essentially three origins: (1) the thermal emission from the wind of the primary, i.e $L_{ff,1}$, (2) the thermal emission from the wind of the secondary, i.e $L_{ff,2}$, and (3) the non-thermal emission from the collision zone between the two stars, i.e $L_{synch}$. This third part is likely to be embedded inside the winds and therefore undergoes a free-free absorption characterized by a given optical depth ($\tau$). We note that a thermal radio emission component is also expected to be produced by colliding winds \citep{Pit2}, but we neglect it in our simplified approach. Consequently, the observed radio luminosity can be expressed as follows:

\begin{equation}
\label{radiotot}
L_{radio}^{obs}\,=\,L_{ff,1} + L_{ff,2} + L_{synch}\,e^{-\tau}
\end{equation}

In order to estimate the non-thermal contribution to the radio spectrum, we need to estimate the free-free thermal contributions from the primary and the secondary respectively. These quantities can be calculated on the basis of the relations given for example by \citet{leietal}, according to the free-free radio emission theory established by \citet{WB} and \citet{PF}:

\begin{equation}
\label{radff}
S_{\nu}\,=\,2.32\,\times\,10^{4}\,\Big(\frac{\dot M\,Z}{V_\infty\,\mu}\Big)^{4/3}\,\Big(\frac{\gamma\,g_\nu\,\nu}{d^3}\Big)^{2/3}
\end{equation}
where $g_\nu$ is the Gaunt factor obtained through the following relation:
$$
g_\nu\,=\,9.77\,\Big(1 + 0.13\,\log\,\frac{T_e^{3/2}}{Z\,\nu}\Big)
$$
The various quantities used in these two relations are
\begin{enumerate}
\item[] $\dot M$ : the mass loss rate in M$_\odot$\,yr$^{-1}$
\item[]	V$_\infty$ : the terminal velocity in km\,s$^{-1}$
\item[] $\nu$ : the frequency in Hz
\item[] $d$ : the distance to the star in kpc
\item[] $T_e$ : the electron temperature, which is generally a fraction of the effective temperature, e.g. $T_e$\,=\,0.5\,$\times$\,$T_{eff}$
\item[] $\mu$ : the mean molecular weight, i.e. $\frac{\sum\,A_i\,M_i}{\sum\,A_i}$, where $A_i$ and $M_i$ are respectively the abundance and the molecular weight of the $i$th ionic species. For a plasma containing 77\,\% hydrogen and 23\,\% helium, $\mu$\,=\,1.69
\item[] $Z$ : the rms ionic charge, i.e. $\frac{(\sum\,A_i\,Z_i^2)^{1/2}}{\sum\,A_i}$, where $Z_i$ is the ionic charge of the $i$th ionic species. For the same abundances as above, $Z$\,=\,1.3
\item[] $\gamma$ : the mean number of electrons per ion, i.e. $\frac{\sum\,A_i\,Z_i}{\sum\,A_i}$. For the same abundances as above, $\gamma$\,=\,1.23
\end{enumerate}

Following this approach, the thermal flux density from both components of the binary system can be calculated. Using Eq.\,\ref{lradio} with a spectral index $\alpha$ of 0.6, these quantities can be converted into radio luminosities expressed in erg\,s$^{-1}$.

Applying Eq.\,\ref{lradio} using $S_{\nu'}$ and $\alpha$ obtained from the observation, the observed radio luminosity can be calculated as well. The difference between these observed (mixed thermal and non-thermal) and theoretical (thermal) quantities provides a lower limit on the luminosity ($L_{synch,min}$) of the intrinsic synchrotron emission produced at the collision zone:
\begin{equation}
\label{lsynchmin}
L_{synch,min}\,=\,L_{synch}\,e^{-\tau}
\end{equation}
where $\tau$ is an optical depth characteristic of the total absorption undergone by the synchrotron radiation produced globally in the emission region located at the collision zone. This optical depth results from the integration of the opacity along the line of sight over the entire emitting volume. The determination of such a quantity is not straightforward and requires an accurate knowledge of the geometrical and wind parameters of the system. This absorption factor ($e^{-\tau}$) is likely to have a large impact on the synchrotron luminosity, mostly in the cases of close binary systems and systems including stars with dense stellar winds (e.g. WR stars). For instance, an optical depth $\tau$ of about 2.3 would be responsible for an attenuation of the synchrotron luminosity by a factor 10. Moreover, we note that since we neglect here the Razin effect (see Sect.\,\ref{synchradproc}) the intrinsic synchrotron luminosity will be underestimated.

\subsection{Estimate of the lower limit on the inverse Compton luminosity}

Provided that (1) the bolometric luminosities of the stars of the binary system are known, (2) the distances between the stars and the collision zone have been calculated, and (3) the lower limit on the intrinsic synchrotron luminosity has been estimated, Eq.\,\ref{lic} can be applied. It should be kept in mind that the intrinsic synchrotron luminosity is expected to vary with the orbital phase if the binary system is eccentric. This comes from the fact that the emitted synchrotron power depends on the strength of the local magnetic field and on the physical extension of the synchrotron emitting region, both of them varying with the orbital phase. When using the minimum synchrotron luminosity as defined by Eq.\,\ref{lsynchmin} at a given orbital phase, this approach leads to a lower limit on the high-energy IC luminosity at the same orbital phase.

\subsection{Estimate of the high-energy cutoff of the non-thermal spectrum\label{heco}}
The electrons are accelerated through the first-order Fermi mechanism, and a valuable quantity worth to be estimated is the maximum energy reached by the electrons under given circumstances. As non-thermal radiation processes provide the most effective cooling mechanisms of relativistic electrons, a high-energy cutoff is obtained when the rate of energy gain of the particles is balanced by the rate of energy loss by radiative processes like IC scattering and synchrotron emission. I note that relativistic electrons may also lose energy through Coulomb collisions with thermal ions, but this cooling process is not expected to be significant \citep{Pit2}: I thus decided to neglect it.

On the one hand, we can use the relation given by \citet{Doug} for the rate of IC loss of energy of relativistic electrons, e.g. electrons whose Lorentz factor $\gamma$ is $\gg$\,1, and the relation provided by \citet{RL} for the energy loss by synchrotron emission:
\begin{equation}
\label{icloss}
\Big.\frac{d\,E}{d\,t}\Big|_{IC}\,=\,\frac{\sigma_T\,\gamma^2}{3\,\pi}\,\frac{L_{bol}}{r^2}
\end{equation}
\begin{equation}
\label{synchloss}
\Big.\frac{d\,E}{d\,t}\Big|_{synch}\,=\,\frac{\sigma_T\,\gamma^2}{6\,\pi}\,c\,B^2
\end{equation}
where $\sigma_T$ is the Thomson cross section whose value is 6.6524\,$\times$\,10$^{-25}$\,cm$^{-2}$. $L_{bol}$ and $r$ are respectively the stellar bolometric luminosity and the distance from the star to the shock region. To take into account the radiation from the two stars of the binary system, the last factor of Eq.\,\ref{icloss} should be written $(\frac{L_{bol,1}}{r_1^2} + \frac{L_{bol,2}}{r_2^2})$.

On the other hand, the rate of energy gain through the Fermi acceleration mechanism given by \citet{baring} can be expressed in cgs units as follows:
\begin{equation}
\label{accgain}
\Big.\frac{d\,E}{d\,t}\Big|_{Fermi}\,=\,A\,\frac{\chi - 1}{\chi\,(1 + g\,\chi)}\,\frac{Q}{\theta}\,B\,V_{sh}^2
\end{equation}
where $A$ is a constant $\sim$\,1.6\,$\times$\,10$^{-20}$.

In this relation, $\chi$ is the compression ratio of the shock, $B$ is the local magnetic field in Gauss, $V_{sh}$ is the speed of the shock in cm\,s$^{-1}$, $g$ is the ratio between the upstream and downstream diffusion coefficients of the high-energy particles generally considered equal to 1/$\chi$, and $\theta$ is the ratio of the mean free path of the particles over the gyroradius. This latter parameter is considered to be constant across the acceleration zone, to be independent of the particle and energy, and to characterize the turbulence. Its minimum value is one (for the case of strong hydromagnetic turbulence), and it was considered to be about 3 by \citet{EU} in the case of the wind collision zone of WR + O binary systems. A low $\theta$ value means that the turbulence is high enough to reduce substantially the mean free path of the particles, therefore confining them in the acceleration region. $Q$ is the mean electric charge.

The particles are accelerated as far as the rate of energy gain is not overwhelmed by the rate of energy loss. The maximum Lorentz factor of the relativistic electrons is consequently provided by the condition:
$$
\Big.\frac{d\,E}{d\,t}\Big|_{IC} + \Big.\frac{d\,E}{d\,t}\Big|_{synch}\,=\,\Big.\frac{d\,E}{d\,t}\Big|_{Fermi}
$$
which translates to
\begin{equation}
\label{gammax}
\gamma^2\,\leq\,\frac{3\,\pi\,A\,(\chi - 1)\,Q\,B\,V_{sh}^2}{\sigma_T\,\chi\,(1 + g\,\chi)\,\theta\,\Big(\frac{L_{bol,1}}{r_1^2} + \frac{L_{bol,2}}{r_2^2} + \frac{c\,B^2}{2}\Big)}
\end{equation}
The maximum Lorentz factor ($\gamma_{max}$) expressed by the right hand side of Eq.\,\ref{gammax} corresponds to an electron energy of $\gamma\,m_e\,c^2$, where $m_e$ is the mass of the electron. Lorentz factors of the order of 10$^{4}$-10$^{5}$ can be reached depending on the long period binary system, with longer periods leading to higher Lorentz factors. For a $\gamma_{max}$ of the order or 10$^{4}$, the maximum energy reached by the electrons is about 5\,GeV.
When the maximum Lorentz factor is estimated, one can determine the maximum frequency of the photons produced by inverse Compton scattering, through the following relation (see Eq.\,\ref{convic}):
\begin{equation}
\label{icmax}
\nu_{max}\,=\,\frac{4}{3}\,\gamma^2_{max}\,\nu_*
\end{equation}
where $\nu_*$ is the typical frequency of the seed stellar photons (see \citet{BluGou}). This quantity can be estimated from Wien's law providing the frequency of the maximum of Planck's function for a black body temperature corresponding to the effective temperature of the star. For massive stars with a typical temperature of about 40000\,K, the radiation field peaks at about 4\,$\times$\,10$^{15}$\,Hz, leading to a maximum energy for the Comptonized photons of the order of a few GeV.\\

In Eq.\,\ref{gammax}, both inverse Compton scattering and synchrotron emission are considered. However, it should be noted that in most cases only IC scattering will significantly contribute to the energy loss of the relativistic particles, and the third term between brackets in the denominator of Eq.\,\ref{gammax} can be neglected. The reason is that the UV and visible radiation fields from the stars are strong, and that the synchrotron mechanism would require too high a local magnetic field strength to become significant.

\section{The multiplicity of non-thermal radio emitters}

As mentioned in Sect.\,\ref{accphys}, the binary scenario appears to be the most adapted to explain the non-thermal radio emission from massive stars. This statement comes mainly from the fact that the binary fraction among these particular sources of radiation is significantly large.

First, let us consider WR stars. A first census of the non-thermal emitting WR stars was presented by \citet{DW}, who pointed out the fact that a large fraction of them were binary systems. At the time of the writing of this paper, 17 WR stars have shown a non-thermal contribution in their radio spectrum  (see Table\,\ref{WRlist}). In most cases, the non-thermal nature of the radio emission is revealed by the spectral index that deviates significantly from 0.6. In some cases, the non-thermal emission is strongly suggested by significant variations in the radio flux. A striking feature of the list provided in Table\,\ref{WRlist} is the large fraction of double or multiple systems among the non-thermally emitting WR stars. Out of 17 non-thermal radio emitters, 14 are at least suspected binaries, with most of them (12) being confirmed binaries. 

In the case of O-type stars, the situation is rather similar. Up to now, 16 non-thermal radio emitters of spectral type O are known (see Table\,\ref{Olist}). In this category, 11 are confirmed binaries and two more are suspected binaries. However, as the situation is less clear than for WR stars, the references for their mutiplicity is also given in Table\,\ref{Olist}.

In order to evaluate the significance of the binary scenario for non-thermal radio emitters, it is useful to give a few words on the binary fraction of massive stars in general. As most massive stars are members of open clusters, an estimate of their binary fraction can be obtained through multiplicity studies of massive stars in open clusters. Some results have been published by \citet{garciamerm}, resulting in intra-cluster binary fractions ranging between 20 and 80\,\% depending on the open cluster. However, in-depth studies of the clusters with the largest claimed binary fractions showed that values of 80\,\% are heavily overestimated. \citet{sana-6231} derived indeed a binary fraction of the order of 60\,\% in NGC\,6231, and \citet{ic1805_2} showed that the binary fraction among massive stars in IC\,1805 should range between 20\,\% and 60\,\%. Putting these numbers with results related to other clusters, the binary fraction among massive stars in general should not be higher than 50\,\% -- and this number should be considered as a very conservative upper limit. The lower limit on the binary fraction among non-thermal radio emitting massive stars is therefore significantly higher than the upper limit on the binary fraction of massive stars in general. Studies are however currently being carried out in order address in detail the issue of the multiplicity of massive stars in general, and of non-thermal radio emitters in particular.

\begin{table}[hb]
\begin{center}
\caption{List of the non-thermal radio emitting WR stars as of November 2005. The information collected in this table is taken from (a) \citet{DW}, (b) \citet{Chap}, (c) \citet{BRKP}, (d) \citet{Cap} and (e) \citet{Mon}. \label{WRlist}}
\vspace*{2mm}
\begin{tabular}{llll}
\hline\hline
Star & Sp. Type & Ref. & Multiplicity \\
\hline
WR~11 & WC8 + O8.5 & a & Confirmed binary with a period of 78.5\,d \\
WR~14 & WC7 & b & Suspected binary \\
WR~21a & WN6 + O3 & c & Confirmed binary \\
WR~39 & WC7 & b & Possible visual binary \\
WR~48 & WC6 + O9.5 & a & Confirmed binary; suspected triple system \\
WR~79a & WN9ha + ? & d & Visual binary \\
WR~89 & WN8ha + OB & d & Visual binary \\
WR~90 & WC7 & b & No evidence for binarity \\
WR~98a & WR + OB & e & Confirmed binary with a period of \\
&  &  & about 1\,yr \\
WR~104 & WR + OB & e & Confirmed binary with a period of \\
&  &  & about 1\,yr \\
WR~105 & WN9 & a & No evidence for binarity \\
WR~112 & WC9 & a & No evidence for binarity \\
WR~125 & WC7 + O9 & a & Confirmed binary with a period of $>$\,15\,yr \\
WR~137 & WC7 + OB & a & Confirmed binary with a period of $>$\,13\,yr \\
WR~140 & WC7 + O5 & a & Confirmed binary with a period of \\
&  &  & about 7.9\,yr \\
WR~146 & WC5 + O8 & a & Visual binary \\
WR~147 & WN8 + B0.5 & a & Visual binary \\
\hline
\end{tabular}
\end{center}
\end{table}

\begin{table}[hb]
\begin{center}
\caption{List of the non-thermal radio emitting O-type stars. The information on the non-thermal nature of the radio emission from the targets collected in this table are taken from (a) \citet{contr}, (b) \citet{BAC}, (c) \citet{setia}, (d) \citet{drake}, (e) \citet{BK}, (f) \citet{leietal}, (g) \citet{BCK}, (h) \citet{Blo167971} and (i) \citet{Blo168112}. For the multiplicity and/or the spectral types, the information are taken from (j) \citet{rauwcyg5}, (k) \citet{Let8a}, (l) \citet{multjenam}, (m) \citet{gies15mon}, (n) \citet{deloritriple}, (o) \citet{ic1805_2}, (p) \citet{nelan}, (q) \citet{rauw93250}, (r) \citet{gies124314}, (s) \citet{gos150136} (t) \citet{Lei87}, and (u) \citet{DeB168112}.\label{Olist}}
\vspace*{2mm}
\begin{tabular}{llll}
\hline\hline
Star & Sp. Type & Ref. & Multiplicity \\
\hline
Cyg\,OB2\,\#5 & O6-7Ib + Ofpe/WN9 & a & Confirmed binary, probably\\
  &  &  & triple$^{j,a}$ \\
Cyg\,OB2\,\#8A & O6If + O5.5III(f) & b & Confirmed binary with a \\
  &  &  & period of 21.908\,d$^k$ \\
Cyg\,OB2\,\#9 & O5If & b & Multiplicity currently \\
  &  &  & under investigation\\
Cyg\,OB2\,-\,335 & O7V & c & Multiplicity not yet \\
  &  &  & investigated \\
9\,Sgr & O4f$^+$ + ? & b & Confirmed binary with a \\ 
  &  &  & period of a few years$^l$ \\
15\,Mon & O7Ve + ? & d & Confirmed binary$^m$ \\
$\delta$\,Ori\,A & O9.5II + ? & b & Confirmed multiple system$^n$ \\
$\sigma$\,Ori\,AB & O9.5V + ? & d & Confirmed multiple system$^d$ \\
HD\,15558 & O5III + O7V & b & Confirmed binary with \\
  &  &  & a period of about 440\,d, \\
  &  &  & possibly triple$^o$ \\
HD\,93129A & O2If + ? & e & Visual binary$^p$ \\
HD\,93250  & O3V((f)) & f & Suspected binary$^q$ \\
HD\,124314 & O6V(n)((f)) + ? & g & Confirmed binary$^r$ \\
HD\,150136 & O5IIIn(f) + ? & g & Confirmed binary, probably \\ 
  &  &  & triple$^s$ \\
HD\,167971 & O5-8V + O5-8V + O8I & b,h & Confirmed binary, \\ 
  &  &  &  probably triple$^t$ \\
HD\,168112 & O5III(f$^+$) & b,i & Suspected binary$^u$ \\
CD\,-47$^\circ$4551 & O5If & g & Multiplicity not yet \\
  &  &  & investigated \\
\hline
\end{tabular}
\end{center}
\end{table}

In the case of B-type stars, only a few objects have been proposed to be non-thermal radio emitters. However, the non-thermal nature of at least a fraction of their radio emission is uncertain and still needs to be confirmed. Considering, the lack of relevant observational indications for a non-thermal radio emission from B stars, this category will not be considered here.\\

The issue of the binarity of many of the objects listed in Tables \ref{WRlist} and \ref{Olist} has been elucidated thanks to spectroscopic observations in the optical domain. In some cases, the multiplicity was studied following high angular resolution techniques. However, our knowledge of the multiplicity of these stars is not complete. In some cases, the multiplicity has not yet been investigated, and in several cases the presence of the companion has only been detected and a lot of information is still lacking concerning the members of these multiple systems. One of the main problems relevant to the issue of the multiplicity of stars in general is the strong bias that affects the census of known binaries. This problem is twofold: (1) A first origin of this observational bias is the problem of the orbital period: typical observing runs last a few days, and are often separated by several months or years. A consequence is that most of the known binaries have periods of a few days, and a few others are just referred to as long period binary candidates because long term radial velocity shifts were detected on time-scales of one or several years. That is the reason why we have only little information on the multiplicity of several targets discussed here. It appears that intensive spectroscopic monitoring should be performed to investigate poorly explored regions of the orbital parameter space, in order to probe some of the fundamental properties of massive stars. Moreover, some of the systems quoted in Table\,\ref{Olist} appear to be multiple systems whose spectroscopic investigation is not straightforward, pointing out the necessity to develop or improve spectral disentangling tools. (2) A second origin of this observational bias is the inclination of the systems we are talking about. As most of the binaries are revealed by spectroscopic monitorings aiming at the measurement of radial velocity excursions, systems characterized by low inclinations are often erroneously classified as single stars. In this context, a promising technique likely to provide a wealth of information is interferometry. Thanks to fairly long baseline interferometers such as the Very Large Telescope Interferometer (VLTI, in Cerro Paranal, Chile) with near infrared spectro-imaging capability, we may for instance confirm the binary status of some stars, and study the orbital motion of long period systems whose orbital parameters have not yet been investigated. Considering the current angular resolution achievable with such interferometers (of the order of a few mas), interferometry should be devoted to multiple systems characterized by rather large separations (at a distance of 1\,kpc, 1\,AU of linear separation corresponds to only about 1\,mas).

The information that can be obtained from such multiplicity studies concern (1) the geometry of the binary or higher multiplicity systems and (2) the spectral and wind properties of their members. Both are crucial in order to constrain the properties of the non-thermal emission components likely to be studied in the radio and high-energy domains (see Sect.\,\ref{hent}).\\

Finally, I should emphasize that the binary/multiple scenario is a valuable working hypothesis which agrees with almost all of the observational constraints available at the time of the writing of this review. More observations are needed to validate -- or invalidate -- this hypothesis. These additional data are mostly required in the case of O-type stars. Indeed, two members of the O-type non-thermal radio emitters catalogue presented in Table\,\ref{Olist} are only suspected binaries (i.e. HD\,168112 and HD\,93250), and recent spectroscopic observations failed to reveal the presence of a companion (see respectively \citet{multjenam} and \citet{audrey}). In other cases, such a spectroscopic monitoring is still lacking. If these stars -- and other non-thermal radio emitters -- are indeed single stars, an alternative scenario will be required to complement the one that is described here. However, as the general scenario described in the previous sections in the context of binary systems still seems to be the most appropriate, such an alternative scenario will not be considered here.

\section{Multiwavelength study of the non-thermal emission\label{obsres}}

Considering the detailed discussion above, we see that a lot of physical aspects should be taken into account simultaneously in order to address the issue of the non-thermal emission from massive binaries in various domains of the electromagnetic spectrum. For this reason, a multiwavelength approach is definitely needed. In this section, I describe several results obtained in various energy domains, and I discuss some prospects for future observations.

\subsection{The radio domain\label{radiodomain}}
As mentioned in the introduction, the non-thermal emission from massive stars was first reported in the radio domain. Many radio observations contributed to the census of non-thermal radio emitters that was made by the end of the 80's \citep{WB,ABC,ATBC,BAC}. In the large radio survey published by \citet{BAC}, about 25\,\% of the early-type stars detected in the radio domain presented the signature of a non-thermal radio spectrum. Since then, this fraction has increased up to a value close to 40\,\%. The present census of non-thermal radio emitters is described in Tables \ref{WRlist} and \ref{Olist} respectively for WR and O-type stars. 

Beside multifrequency radio observations of massive stars able to reveal flux densities and spectral indices, crucial information came from high-angular resolution radio observations (VLA, VLBA, MERLIN). Thanks to these data, major progres has been made by resolving non-thermal emission regions in the case of a few colliding-wind binaries, therefore confirming that the non-thermal emission arises from the wind-wind interaction region and not from the stars themselves. Among these examples, we may consider Cyg\,OB2\,\#5 \citep{contr}, WR\,146 \citep{oconnorwr146}, WR\,147 \citep{williamswr147} and WR\,140 \citep{Doug140}.\\

An important point worth mentioning here concerns the confrontation of observational radio results with detailed simulations of radio emission from single O-stars. \citet{sven2006} demonstrated that the observed non-thermal radio emission from O-stars cannot be explained by current models of wind-embedded shocks produced by the instability of the line-driving mechanism, therefore lending strong support to the binary scenario for the non-thermal radio emission from massive stars. This latter scenario is thus supported by both theoretical (radio emission models) and observational (multiplicity studies) arguments.

\subsection{The soft X-ray domain}

One of the main questions that has been addressed in the context of ongoing multiwavelength observations is that of the high-energy counterpart to the non-thermal emission already detected in the radio domain. As shown in Sect.\,\ref{radphys}, several processes may be responsible for the production of a non-thermal high-energy emission. Among the processes discussed in Sect.\,\ref{radphys}, inverse Compton scattering is expected to be the most significant one as it requires a large amount of UV and visible photons, and massive stars are strong sources over precisely these energy domains. As a consequence, we may expect a non-thermal component in the high-energy domain following a power law spectral shape:
\begin{equation}
F(E) \propto E^\alpha
\end{equation}
The spectral index ($\alpha$) is directly related to the index of the relativistic electrons:
\begin{equation}
\alpha = \frac{n - 1}{2}
\end{equation}
As a result, in the case of strong shocks where $n$ is equal to 2 (see Eq.\,\ref{fermiindex}), we expect the non-thermal emission component in the high-energy domain to have a spectral index of 0.5.

Several non-thermal radio emitting massive stars have been observed with the European X-ray satellite {\it XMM-Newton} that is sensitive at energies between 0.5 and 10.0\,keV. As mentioned in Sect.\,\ref{windphys}, massive stars and massive binaries are thermal emitters in the X-ray domain. These thermal emission components are characterized by plasma temperatures intimately dependent on the pre-shock velocities of the plasmas that collide \citep{SBP}. In the case of intrinsic shocks in individual stellar winds, the pre-shock velocities are of the order of a few 100\,km\,s$^{-1}$. The post-shock temperature is therefore of the order of a few 10$^{6}$\,K. In the case of colliding winds, the situation is a bit more complicated. If we consider long period binaries, i.e.\,binaries with orbital periods longer than a few weeks, the separation between the stars in the binary system is rather large. As a consequence, the winds have enough space (and time) to accelerate up to their asymptotic velocity (the terminal velocity) before colliding. In such a situation, the typical pre-shock velocity is of the order of 2000\,km\,s$^{-1}$ and the post-shock temperature is of the order of several 10$^{7}$\,K. In the case of close binaries, i.e.\,binaries with periods not longer than a few days, the winds collide in their acceleration zone. The pre-shock velocities are therefore much lower than in the case of long period binaries, i.e.\,of the order of 1000\,km\,s$^{-1}$, and the post-shock temperature is of the order of 10$^{7}$\,K.

As thermal processes dominate the X-ray spectrum of massive stars in the {\it XMM-Newton} bandpass, we may expect the thermal emission components to have a strong impact on our capability to detect a non-thermal emission component in the X-ray spectrum of massive binaries: {\it a putative non-thermal emission component will be detected only if it is not overwhelmed by the thermal emission}.\\

Several O-type non-thermal radio emitters have been observed with {\it XMM-Newton}: 9\,Sgr \citep{rauw9sgr}, HD\,168112 \citep{DeB168112}, HD\,167971 \citep{DeB167971}, Cyg\,OB2\,\#8A \citep{DeBcyg8a} and Cyg\,OB2\,\#9 \citep{jenamcygob2}. These targets are confirmed or suspected long period binaries. In each case, we observe an X-ray spectrum characteristic of rather high plasma temperatures. Moreover, the spectrum is hard, i.e.\,the X-ray flux descreases rather smoothly as a function of the energy. None of these targets revealed unambigously a non-thermal X-ray contribution in their X-ray spectrum below 10.0\,keV. In addition, it is worth mentioning that the analysis by \citet{skinner-wr147} of {\it XMM-Newton} data of the very long period WR + O binary WR\,147 reveals a thermal spectrum below 10\,keV. On the other hand, the {\it XMM-Newton} observation of the short period colliding-wind binary HD\,159176 \citep{DeB159176} revealed a rather soft thermal spectrum along with a power law suggesting the presence of a non-thermal emission component. If it is confirmed, HD\,159176 would be the first O-type non-thermal emitter in the X-ray domain below 10.0\,keV. This result is therefore very interesting, mostly if we consider that this latter target is {\it not} a non-thermal radio emitter. A second candidate for non-thermal emission from a short period binary in the soft X-ray domain is FO\,15 \citep{fo15}.\\

\begin{figure}[h]
\centering
\includegraphics[width=11.5cm]{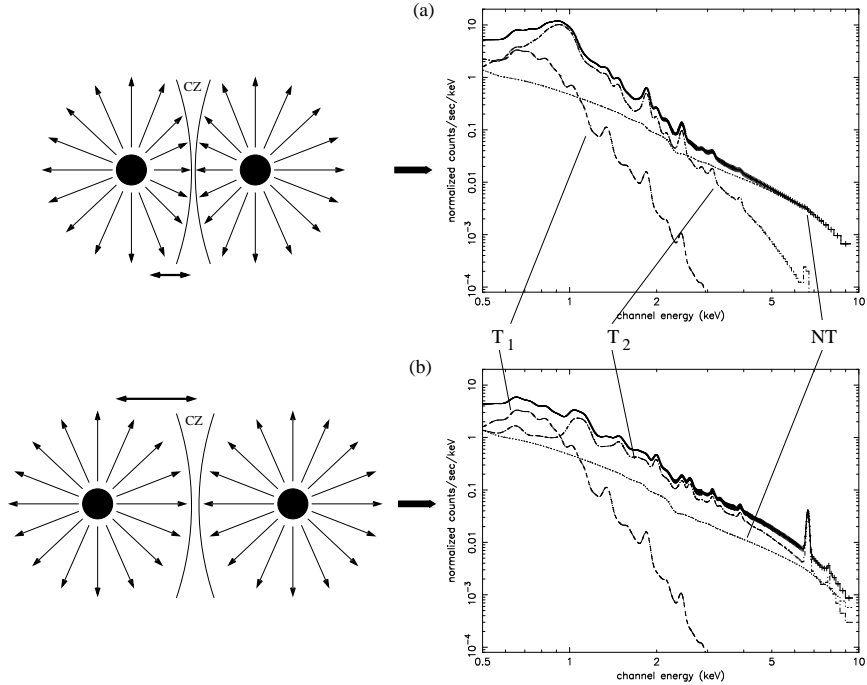}
\caption{Effect of the hardness of the thermal X-ray spectrum on the detectability of a putative non-thermal emission component. Cases (a) and (b) stand respectively for {\it short period} and {\it long period} binaries. On the left: schematic view of the binary system, with the bold double arrows emphasizing the significant difference in the separation between the star and the collision zone (CZ) depending on the case (long or short period). On the right: simulated {\it XMM-Newton} spectrum between 0.5 and 10.0\,keV including thermal (T$_1$ and T$_2$) and non-thermal (NT) components. The temperature of the T$_1$ component and the spectral index of the NT component are the same in the two cases. The only difference resides in the temperature of the second thermal component, which is higher in case (b) than in case (a). In the simulated spectrum of case (b), the strong spectral line at about 6.7\,keV is due to Fe, and is often considered as a signature of a high plasma temperature. Case (a): the power law is revealed, at least above about 3\,keV. Case (b): the power law is overwhelmed by the high temperature thermal component, and is therefore not revealed. \label{hardness}}
\end{figure}

Considering these observational results, we may distinguish two main categories. The first category is that of {\it long period} binaries, i.e.\,binaries with periods longer than a few weeks. In this case, the separation is large and the stellar winds reach their terminal velocities before they collide. The post-shock temperature is high and the thermal emission is hard. In addition, the collision zone is far from the stars and the UV radiation field is weak in the interaction zone. This situation is therefore not favourable to the inverse Compton scattering process that requires a large amount of low energy photons to be efficient. The putative non-thermal emission is thus expected to be weak. As a result, long period binaries are unlikely to reveal any non-thermal contribution in their soft X-ray spectrum (see Fig.\,\ref{hardness}, part (b)). In the radio domain, the large binary separation allows a significant fraction of the synchrotron radio emission to emerge from the combined stellar winds, and such systems may appear as non-thermal radio emitters.

In the case of {\it short period} binaries, i.e.\,binaries with periods not longer than a few days, the binary separation is shorter. As a consequence, the post-shock temperature at the collision zone is lower and the thermal X-ray spectrum is softer, i.e.\,the X-ray flux decreases more rapidly as the energy increases. Moreover, the short distance between the stars and the collision zone is in favour of the inverse Compton scattering mechanism. The relative intensity of the putative non-thermal emission component may be rather high. In this situation, a non-thermal emission component may be revealed in the X-ray domain (see Fig.\,\ref{hardness}, part (a)). In the radio domain, the synchrotron emission is unlikely to be unveiled as the emission region is strongly burried in the combined stellar winds, and therefore significantly absorbed. As a result, short period binaries will not be detected as non-thermal radio emitters. Such systems may be referred to as {\it synchrotron-quiet} non-thermal emitters, provided they produce a detectable non-thermal radiation in the high energy domain but not at radio wavelengths.\\

{\it In summary, it appears that the simultaneous detection of a non-thermal emission in the radio domain and in the X-ray domain below 10.0\,keV is unlikely. However, as the detection in the X-ray domain is mostly inhibited by the presence of the strong thermal emission, a simultaneous detection can be envisaged if we investigate the higher energies -- the hard X-rays (above 10.0\,keV) and the soft $\gamma$-rays -- where the thermal emission is absent} \citep{thesis}.

\subsection{The hard X-ray and soft $\gamma$-ray domains}

The detection of a high-energy counterpart to the non-thermal radio emission still needs to be achieved in the case of O-type stars. The analysis of {\it INTEGRAL}-ISGRI data performed by \citet{cygint} between 20\,keV and 1\,MeV failed to reveal any hard X-ray of soft $\gamma$-ray emission related to the O-type star population of the Cyg\,OB2 region. 

Another point worth mentioning concerns the magnetic field. As shown in this paper, the magnetic field intervenes in differents aspects of the question of the non-thermal emission from massive binaries (synchrotron radio emission, acceleration of particles...). One can therefore wonder whether the study of the non-thermal emission from massive stars could lead to an indirect method to estimate the magnetic field strength, at least locally, in the collision zone. Indeed, when we consider for instance Eq.\,\ref{lic}, we see that in the case of a binary system, the simultaneous knowledge of several parameters such as the bolometric luminosities, the distance between the stars and the interaction zone, the synchrotron radio luminosity and the high-energy inverse Compton luminosity, may allow to estimate the strength of the local magnetic field. However, as discussed above, the simultaneous determination of the latter two quantities requires the combination of radio and high-energy (hard X-rays and soft $\gamma$-rays) observations. This prospect underlines the interest of observing non-thermal radio emitters at higher energies.

\subsection{The very high-energy $\gamma$-rays}

When we are dealing with very high energy processes, we should keep in mind the fact that energy of the particles involved in these processes must be very high as well (i.e. particles can only transfer to radiation the energy they possess).
As discussed in Sect.\,\ref{heco}, the maximum Lorentz factor that can be reached by relativistic electrons in colliding-wind binaries should be of the order of 10$^4$ to 10$^5$ \citep{thesis,Pit2}. As a result, the high-energy photons produced  by electrons through inverse Compton scattering should not reach energies higher than a few GeV. However, we should not forget that protons are also accelerated. In Eq.\,\ref{accgain}, we see that the maximum Lorentz factor reached by particles depends on their mean free path in the acceleration zone, i.e. the typical length scale between two scattering centers. Generally, it is considered that this mean free path is directly related to the gyroradius of the particle. The more massive the particles (protons are about 1800 times more massive than electrons), the shorter the gyroradius and therefore shorter is their mean free path. The confinement of protons in the acceleration zone will therefore be significantly improved as compared to that of electrons. Protons are thus expected to reach much higher energies than electrons.

Considering the dependence of the Thomson cross section on the mass of the particle (see \citet{Long}), inverse Compton scattering by protons will be inefficient. As a result, other hadronic processes\footnote{Hadronic processes are processes involving hadrons (i.e. protons, neutrons or ions), in contrast to leptonic processes requiring the intervention of leptons such as electrons.} such as neutral pion decay should be considered (see Sect.\,\ref{radphys}). This latter process has indeed been proposed to be at work in the vicinity of massive stars \citep{CWBpi,DSACWB,PD140art}.

With the advent of ground-based Cherenkov telescopes able to detect $\gamma$-ray photons with energies between about 100\,GeV and a few TeV, a new window of high-energy astrophysics is now open. Among the so-called TeV sources that have been detected in the recent years, some of them may be related to early-type star populations. The first unidentified TeV source was indeed discovered in the rich OB association Cyg\,OB2 by the High Energy Gamma Ray Astronomy consortium \citep{Tev}, and its position has been slightly revised recently thanks to observations with the Whipple Observatory 10\,m telescope \citep{tev-konopelko}. The nature of this source is still unknown, even though it has been proposed that it may be related to the rich population of massive stars in the Cyg\,OB2 region \citep{buttcwe-tev,torres-assoc}. It is worth noting that the recent study by \citet{butt-tev} revealed the existence of a radio counterpart to this TeV source. Another TeV source was discovered close to the Galactic Center using the same technique \citep{TevGC}. This latter source was proposed by \citet{QL} to be related to the massive star population located in the central region of the Galaxy. However, a detailed study of the massive star population close to the Galactic Center is still lacking, although some recent investigations showed that tens of early-type stars (O-type and WR) are present in that region \citep{paumGC}.  

\section{Recent theoretical developments\label{theordev}}

The many observational results gathered concerning the non-thermal radio emission from massive binaries, along with the prospects for the detection of a high-energy counterpart, motivated astrophysicists to develop models devoted to this issue. Significant advances have been made during the last years.

In the radio domain, recent models by \citet{Doug,Dougjenam} improved significantly our vision of the physical processes at work in colliding-wind binaries. These new approaches use hydrodynamical codes to describe the stellar winds and the wind collision regions in order to model the synchrotron emission, taking into account several crucial effects such as the Razin effect or the free-free absorption by the wind material. More recently, the detailed work by \citet{Pit2} improved significantly the modelling of non-thermal emitting colliding winds through the inclusion of the inverse Compton cooling. The strong impact of this latter process is indeed likely to alter substantially the properties of the population of relativistic electrons, and therefore the synchrotron emission as well. These models predict also a steepening of the electron spectrum, i.e. an electron index larger than 2 (see Sect.\,\ref{accphys}), where IC cooling is strong. This comes from the fact that IC cooling appears to be more efficient at cooling the higher energy electrons. It is interesting to note that this may explain why the spectral index detected in the case of HD\,159176 -- the first O-type binary possibly presenting a non-thermal component in the soft X-rays -- is significantly larger than 0.5 which is the expected theoretical value for strong shocks. The inclusion of IC cooling allows moreover to estimate the non-thermal high-energy spectrum of colliding-wind binaries following a more satisfactory approach than that allowed by Eq.\,\ref{lic}. These recent studies allow indeed to fit models to radio data in order to determine the population and spatial distribution of relativistic electrons without an a priori knowledge of the local magnetic field. In addition, it appears that the relativistic particles themselves are expected to alter the hydrodynamic properties of the shocks responsible for the particle acceleration, like in the case of supernova remnants (see e.g. \citet{Decourchelle}). When this {\it shock modification} is efficient, the relativistic electron population is characterized by an index (see Eq.\,\ref{fermipower}) larger that 2 (see \citet{PD140art}). On the other hand, \citet{PD140art} suggested that in the case of WR\,140 (see below), the energy spectrum of the relativistic electrons may be affected by the multiple accelerations likely to occur consecutively in different shocks. In this scenario, the electrons are first accelerated by intrinsic shocks in individual stellar winds, before being further accelerated by the shocks of the wind-wind interaction region. When this {\it re-acceleration} process occurs, the spectral index of the relativistic electrons is lower than 2 \citep{PD140art}.  

Alternative models have also been developed by \citet{DSACWB} in order to address the issue of the non-thermal high-energy emission from colliding-wind binaries. This latter approach includes the particle acceleration self-consistently and allows to compute high-energy spectra through various emission processes such as inverse Compton scattering (in the Thomson and Klein-Nishina regimes) or neutral pion decay. Such models are necessary in order to prepare future high-energy observations with space-borne observatories such as {\it GLAST}, or ground-based Imaging Atmospheric Cherenkov Telescopes (IACTs).

\section{Two particularly interesting systems: WR\,140 and Cyg\,OB2\,\#8A}
In this section, I discuss more specifically the case of two non-thermal radio emitters that deserve particular attention. These two cases have been selected to span a rather large domain of the stellar and orbital parameter space. The first one is a long period binary system made of a WR and of an O-type star (WR\,140), and the second one is a shorter period binary constituted of two O-type stars (Cyg\,OB2\,\#8A).

\subsection{WR\,140}
This interesting binary was the first star for which non-thermal emission was detected \citep{FG140}, and it has long been considered as the archetypal non-thermal emitting WR star. It consists in a WC7 primary with an O4-5 secondary orbiting with a period of about 7.9\,yr in a highly eccentric orbit (e\,=\,0.88). \citet{wil140} showed that the highly elliptical orbit is responsible for strong variations all the way from X-rays to radio wavelengths. The intensive long term radio monitoring of WR\,140 performed by \citet{WhBe140} provided for the first time a radio light curve of a non-thermal radio emitter during a complete orbital cycle. This light curve presented strong modulations unlikely to be attributed to free-free absorption only. The variations are believed to be at least partly due to intrinsic changes in the physical conditions underlying the intrinsic synchrotron emission as a function of the orbital phase. More recently, \citet{Doug140} provided a more detailed view of the dependence of the properties of the synchrotron radio spectrum of WR\,140 as a function of the orbital phase through multifrequency observations. In addition, the latter authors reported on an arc of emission -- reminiscent of the bow-shaped morphology expected for the wind collision region -- that is rotating as the orbit progresses, even though it should be noted that the observed morphology is also affected by the absorption of radiation by the wind material close to the shock.

WR\,140 has been proposed as a possible counterpart to the unidentified EGRET $\gamma$-ray source 3EG\,J2022+4317 \citep{BRwr}, even though the binary lies at the limit of its error box. The investigation of a large set of {\it INTEGRAL}-IBIS data did not lead to the detection of a hard X-ray emission (between 20\,keV and 1\,MeV) from WR\,140 \citep{cygint}. On the basis of models such as discussed in Sect.\,\ref{theordev}, \citet{PD140,PD140art} calculated the expected high-energy spectrum of WR\,140. From the fit of the radio data at a particular orbital phase, these authors estimated that the high-energy emission would be dominated by IC scattering up to 50\,GeV, while the emission from hadronic processes may reach a few TeV. The predicted emission levels suggest promising results from future observations with GLAST and Cherenkov air shower telescope arrays. A future detection of WR\,140 with these latter instruments is also strongly suggested by the results of \citet{DSACWB}. 

\subsection{Cyg\,OB2\,\#8A}
This particularly interesting target was unambiguously classified as a non-thermal radio emitter by \citet{BAC}. The spectroscopic study of \citet{Let8a} led to the discovery of its binarity, with an orbital period of about 22 days, and \citet{DRcanada} provided some indications of a wind-wind interaction in this system. In the X-rays below 10\,keV, the studies performed for instance by \citet{master} and \citet{DeBcyg8a} did not reveal any non-thermal emission component. The soft X-ray spectrum appears indeed to be essentially dominated by the strong thermal and phase-locked variable emission from the colliding-winds \citep{jenam8a,DeBcyg8a}. In the radio domain, the recent monitoring performed by \citet{Blomme8a} revealed a strong phase-locked variability of the radio flux, in excellent agreement with the ephemeris published by \citet{Let8a}. As the radius of the sphere with unit optical depth in the radio domain is expected to be very large (of the order of 100\,R$_\star$)\footnote{The radius of the sphere with unit radio optical depth is a decreasing function of the frequency. It may be of the order of a few tens of stellar radii at 3\,cm, and it can reach several hundreds of stellar radii at 20\,cm.}, it was a priori unexpected to detect the synchrotron radio emission that is most probably produced in the wind interaction region, i.e. deep inside the combined winds. The fact that we clearly detect it is most probably due to a combination of wind porosity and inclination effects. A more detailed monitoring of the orbital cycle in the radio domain is planned in the future in order to constrain the physical cirucumstances underlying the non-thermal emission processes at work in the colliding-wind region.

A point that deserves to be mentioned is that Cyg\,OB2\,\#8A -- along with other non-thermal radio emitting massive stars -- lies inside the error box of the unidentified EGRET source 3EG\,2032+4118. This $\gamma$-ray source has already been proposed to be associated to massive stars in the Cyg\,OB2 association: Cyg\,OB2\,\#5  \citep{BR}, Cyg\,OB2\,\#8A \citep{HK}. However, recent observations with the IBIS imager onboard the European {\it INTEGRAL} satellite between 20\,keV and 1\,MeV did not allow to detect the EGRET source \citep{cygint}. Observations with future hard X-ray observatories benefitting from a better sensitivity than {\it INTEGRAL} are required to address the issue of the non-thermal high-energy emission from this promising system. In this context, observations between 10 and 100\,keV with the SIMBOL-X satellite are very promising \citep{debecker-simbolx}.

Among the O-type non-thermal radio emitters, Cyg\,OB2\,\#8A appears to be the best candidate for an application of state-of-the-art models \citep{Doug,Pit2}. Cyg\,OB2\,\#8A is indeed the O-type non-thermal radio emitter with the best constrained stellar, wind, and orbital parameters \citep{Let8a,thesis}, and constitutes therefore a privileged target to probe the physics underlying the non-thermal processes in massive colliding-wind binaries.

\section{Concluding remarks}

Since the discovery of the occurrence of non-thermal processes in several massive stars, significant progresses have been achieved. First, the nature of non-thermal radio emission seems to be established: it has been identified as synchrotron radiation. Second, we are more and more convinced that the binary scenario is the most adapted to explain the non-thermal emission observed in the radio domain.

A crucial point that is extensively demonstrated in this review is the necessity to address the issue of the non-thermal emission from massive binaries following a multiwavelength approach. The discussion of the two best studied members of the catalogues presented in Tables \ref{WRlist} and \ref{Olist} -- WR\,140 and Cyg\,OB2\,\#8A -- shows that many types of information on these systems (stellar parameters, wind properties, orbital elements) are required in order to discuss in detail their non-thermal emission in several energy domains, through the application of recent theoretical models. These models are now being applied to a few WR-type non-thermal radio emitters. The next step will consist in their applicaiton to O-type systems such as Cyg\,OB2\,\#8A. Such an extension to O-type binaries is a necessary condition to probe a more extended domain of the parameter space. In addition, the investigation of synchrotron-quiet massive binaries should not be neglected as the latter may also constitute non-thermal radiation sources in the high-energy domain. As a result, it may be more justified to talk about a more extended class of objects called {\it non-thermal emitters}, among which one could find systems detected as non-thermal radio sources -- the so-called non-thermal radio emitters -- and possibly non-thermal emitters in the high-energy domain that are not detected as synchrotron emitters in the radio domain.  

Even though leptonic processes are by far the most studied in this context, the prospect of very high $\gamma$-ray emission from massive stars deserves to consider hadronic processes as well. State-of-the-art models now include both processes simultaneously, and provide powerful tools likely to sketch a more complete view of the multiwavelength spectrum of non-thermal emitting early-type colliding-wind binaries. Additional studies aiming at the improvement of the knowledge of useful parameters are required, notably in the visible and near-infrared domains. Such studies are needed in order to investigate for instance the acceleration of particles in colliding winds of early-type stars, and whether massive binaries may contribute to very high-energy sources such as the TeV emitters that have been detected using Cherenkov telescopes in the Cyg\,OB2 association \citep{Tev} or close to the Galactic Center \citep{TevGC}. 

\begin{acknowledgements}
I wish to acknowledge support from the PRODEX XMM-Newton/INTEGRAL Projects and from the Fonds National de la Recherche Scientifique (FNRS, Belgium). It is a pleasure to thank Dr. Gregor Rauw for many valuable comments on this manuscript, along with Drs. Ronny Blomme and Julian Pittard for collaborations and fruitful discussions related to the topic of this review. Many thanks to the referee whose comments significantly helped to improve this paper. I am also grateful to Profs. Jean-Pierre Swings and Jean Surdej for encouraging me to write a review on this topic, and to the Editorial Committee of Astronomy \& Astrophysics Review for giving me the opportunity to publish it. 
\end{acknowledgements}

\bibliographystyle{spbasic}

\end{document}